\documentclass[10pt,conference]{IEEEtran}

\usepackage{cite}
\usepackage{amsmath,amssymb,amsfonts}
\usepackage{algorithm}
\usepackage{algpseudocode}
\usepackage{graphicx}
\usepackage{textcomp}
\usepackage{xcolor}
\usepackage[hyphens]{url}
\usepackage{fancyhdr}
\usepackage[colorlinks,citecolor=blue!60!black,linkcolor=blue!60!black,urlcolor=blue!60!black]{hyperref}
\usepackage{booktabs}
\usepackage{tabularx}
\usepackage{makecell}
\usepackage{xspace}
\usepackage{tikz}
\usetikzlibrary{patterns, decorations.pathreplacing, arrows.meta, positioning}
\usepackage{microtype}

\title{Dynamic HBM Repartitioning for Multi-Turn MoE Serving}

\providecommand{\Description}[1]{}

\newcommand{\squishlist}{
   \begin{list}{$\bullet$}
    { \setlength{\itemsep}{0pt}      \setlength{\parsep}{0pt}
      \setlength{\topsep}{-3pt}       \setlength{\partopsep}{0pt}
      \setlength{\listparindent}{-2pt}
      \setlength{\itemindent}{-5pt}
      \setlength{\leftmargin}{1em} \setlength{\labelwidth}{0em}
      \setlength{\labelsep}{0.5em} } }

\newcommand{\squishend}{
    \end{list}\vspace{3pt}  }

\makeatletter
\newcommand{\linebreakand}{%
  \end{@IEEEauthorhalign}
  \hfill\mbox{}\par
  \mbox{}\hfill\begin{@IEEEauthorhalign}
}
\makeatother

\author{
  \IEEEauthorblockN{Jinpyo Kim}
  \IEEEauthorblockA{UC San Diego\\
    jik066@ucsd.edu}
  \and
  \IEEEauthorblockN{Mingi Kwon}
  \IEEEauthorblockA{UC San Diego\\
    mik090@ucsd.edu}
  \and
  \IEEEauthorblockN{Younghoon Min}
  \IEEEauthorblockA{SK hynix\\
    younghoon.min@sk.com}
  \linebreakand
  \IEEEauthorblockN{Jongryool Kim}
  \IEEEauthorblockA{SK hynix\\
    jongryool.kim@sk.com}
  \and
  \IEEEauthorblockN{Jishen Zhao}
  \IEEEauthorblockA{UC San Diego\\
    jzhao@ucsd.edu}
}

\begin{document}
\maketitle
\thispagestyle{plain}
\pagestyle{plain}

\newcommand{\sys}{\textsc{VAMP}\xspace}
\newcommand{\sysfull}{\textbf{VAMP}: Value-Aware Memory Partitioning\xspace}
\newcommand{\hone}{\textbf{H100-PCIe}\xspace}
\newcommand{\htwo}{\textbf{H200-NVL}\xspace}
\newcommand{\vllm}{vLLM\xspace}
\newcommand{\swebench}{SWE-bench Lite\xspace}

\newcommand{\figThesis}{%
\begin{figure}[t]
\centering
\includegraphics[width=\linewidth]{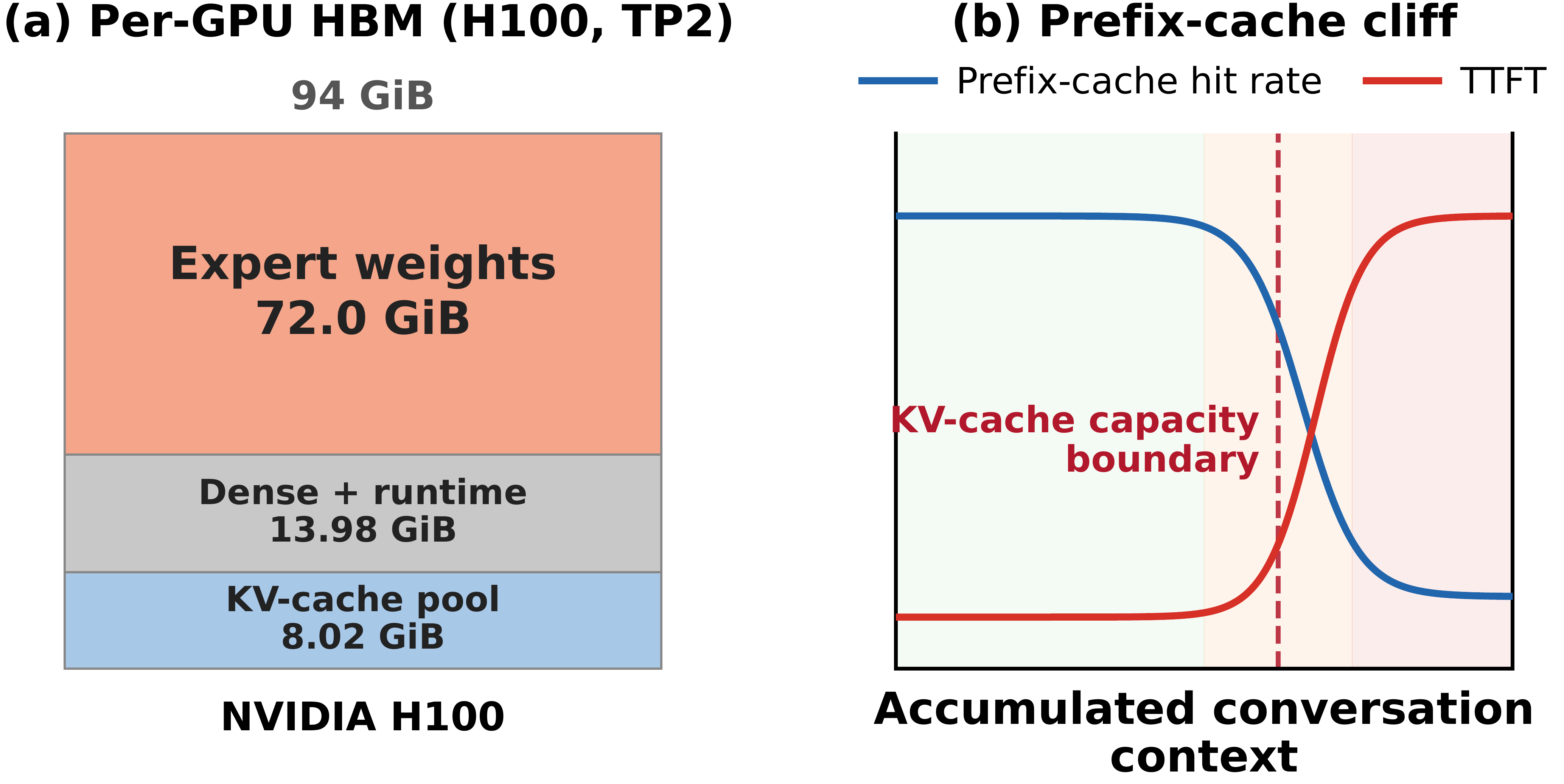}
\caption{\textbf{HBM pressure in MoE serving.} (a) The per-GPU HBM
  quantities reported in Table~\ref{tab:platforms} for the \hone{}
  TP2 configuration: expert weights occupy 72.0\,GiB and the baseline
  KV-cache pool 8.02\,GiB of the 94\,GiB device capacity; the
  \emph{Dense + runtime} region (non-expert model state, runtime
  overhead, and reserved headroom) is the derived, unitemized
  remainder. (b) Schematic illustration
  of the expected prefix-cache cliff: as conversation context
  accumulates across turns and its reusable prefix-cache footprint exceeds the
  fixed KV-cache capacity, cache eviction reduces the prefix-cache hit
  rate and sharply increases TTFT.}
\Description{Two-panel figure. Panel (a), titled per-GPU HBM for H100 under tensor parallelism two, is a block representing 94 GiB of HBM divided into three regions stacked top to bottom: a salmon region labeled expert weights, 72.0 GiB; a gray region labeled dense plus runtime, 13.98 GiB; and a blue region at the base labeled KV-cache pool, 8.02 GiB. Region areas are not drawn to scale; the printed values carry the true proportions. Panel (b), titled prefix-cache cliff, is a line chart with no numeric scales; the horizontal axis is labeled accumulated conversation context, green, orange, and red background bands run from left to right, and a legend names the two curves. A thick blue curve, prefix-cache hit rate, stays high and then falls steeply near a dark-red dashed vertical line labeled KV-cache capacity boundary, while a thick red curve, TTFT, stays low and rises sharply around the same boundary.}
\label{fig:thesis}
\end{figure}%
}

\newcommand{\figMotivation}{%
\begin{figure*}[t]
\centering
\includegraphics[width=\textwidth]{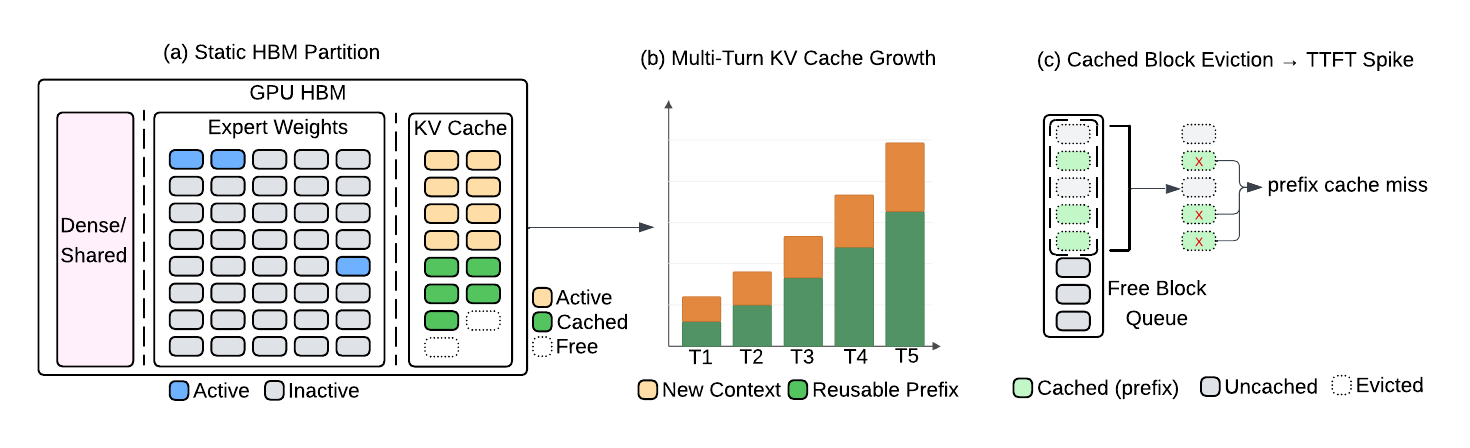}
\caption{Limitations of existing memory management for multi-turn
  MoE serving. (a) Static partitioning keeps inactive expert memory
  from the KV-cache pool. (b) KV-cache footprint and reusable prefix
  both grow across turns. (c) Eviction is cost-blind.}
\Description{Three-panel diagram illustrating limitations of static memory management. Panel (a) shows a GPU HBM bar divided into two fixed regions: a large block for expert weights (with hatching indicating inactive experts) and a small block for the KV-cache pool, with an annotation showing that the inactive expert memory is walled off and unavailable to the KV cache. Panel (b) shows a timeline across conversation turns where two lines, one for KV-cache size and one for reusable prefix length, both increase monotonically across turns, illustrating that agentic sessions continuously accumulate context with no natural release. Panel (c) shows a block-allocation diagram across several turns where a cost-blind LRU policy evicts cached prefix blocks that are subsequently needed by later turns, resulting in re-prefill events and a spike in TTFT indicated by a red annotation.}
\label{fig:motivation}
\end{figure*}%
}

\newcommand{\figExpertOffload}{%
\begin{figure}[t]
  \centering
  \includegraphics[width=0.85\columnwidth]{figures/fig_expert_offload.pdf}
  \caption{Expert offloading: MoE expert weights are offloaded from GPU HBM to CPU DRAM and transferred back over PCIe when needed.}
  \Description{Diagram showing two memory regions connected by a bidirectional arrow. On the left, a box labeled GPU HBM contains several colored blocks representing expert weight tensors, with some blocks faded or hatched to indicate they have been offloaded. On the right, a box labeled CPU DRAM holds a corresponding set of expert weight blocks. The arrow between the two boxes is labeled PCIe, indicating the data transfer path. An annotation clarifies that offloaded experts must be transferred back over PCIe before the corresponding MoE computation can proceed.}
  \label{fig:expert_offload}
\end{figure}%
}

\newcommand{\figKVCacheDialog}{%
\begin{figure*}[t]
  \centering
  \includegraphics[width=.9\textwidth]{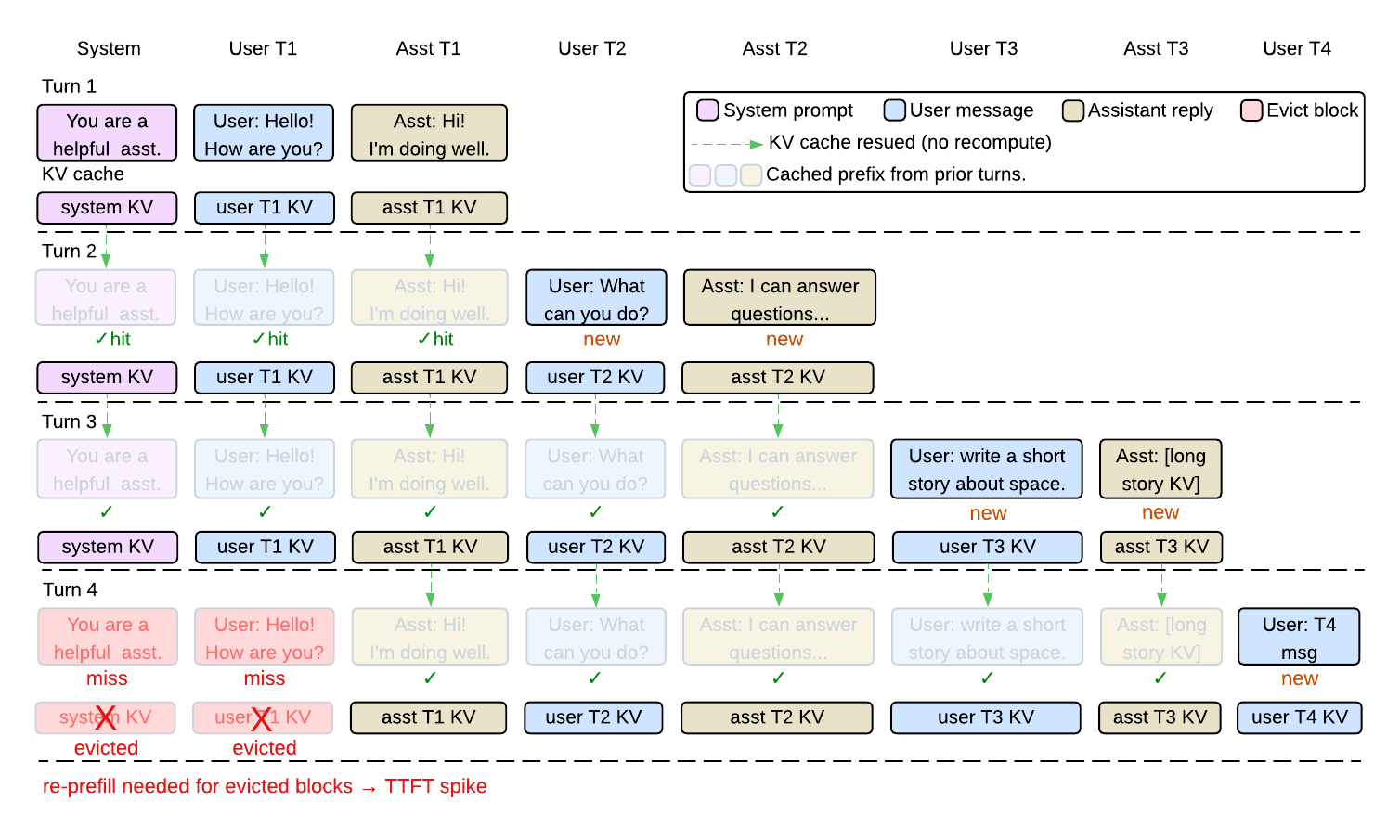}
  \caption{KV cache reuse and eviction across a multi-turn agentic session. Turns 1--3 accumulate cached blocks that are reused (green arrows, no recompute) on subsequent turns. In Turn 4, two early blocks are evicted to free space for the new request; those blocks become cache misses, forcing re-prefill and causing a TTFT spike.}
  \Description{Sequence diagram depicting four turns of a multi-turn agentic session. Each turn is shown as a row of colored blocks representing KV cache entries for that turn's tokens. Turns 1 through 3 show cached blocks accumulating and green arrows pointing from earlier blocks to later turns, indicating successful prefix-cache hits where those blocks are reused without recomputation. In Turn 4, two blocks from earlier in the session are marked with red X symbols indicating eviction, and the subsequent cache-miss events are annotated with a re-prefill label and an upward-pointing arrow labeled TTFT spike, illustrating the cascading latency cost caused by cost-blind eviction.}
  \label{fig:kv_cache_dialog}
\end{figure*}%
}

\newcommand{\figOverview}{%
\begin{figure}[t]
  \centering
  \includegraphics[width=.9\columnwidth]{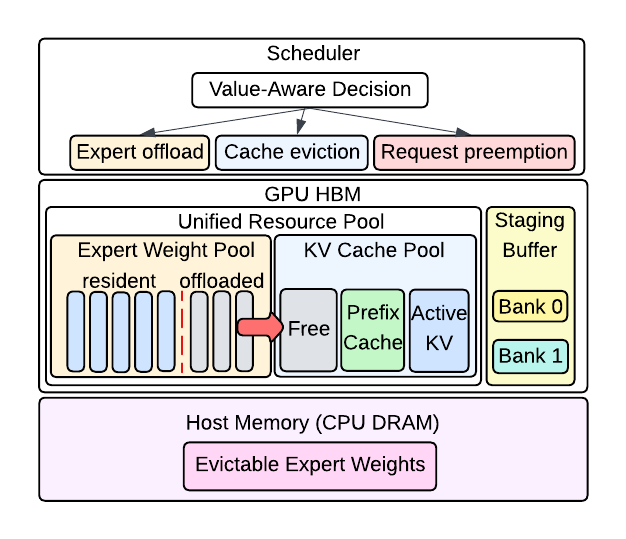}
  \caption{Overview of the \sys architecture.}
  \Description{System overview diagram of VAMP, illustrating how the value-aware action selection, dynamic resource orchestrator, and batching-aware expert offloading components interact to manage HBM as a unified pool shared between expert weights and the KV cache.}
  \label{fig:overview}
\end{figure}%
}

\newcommand{\figLayerDemand}{%
\begin{figure}[t]
    \centering
    \includegraphics[width=\columnwidth]{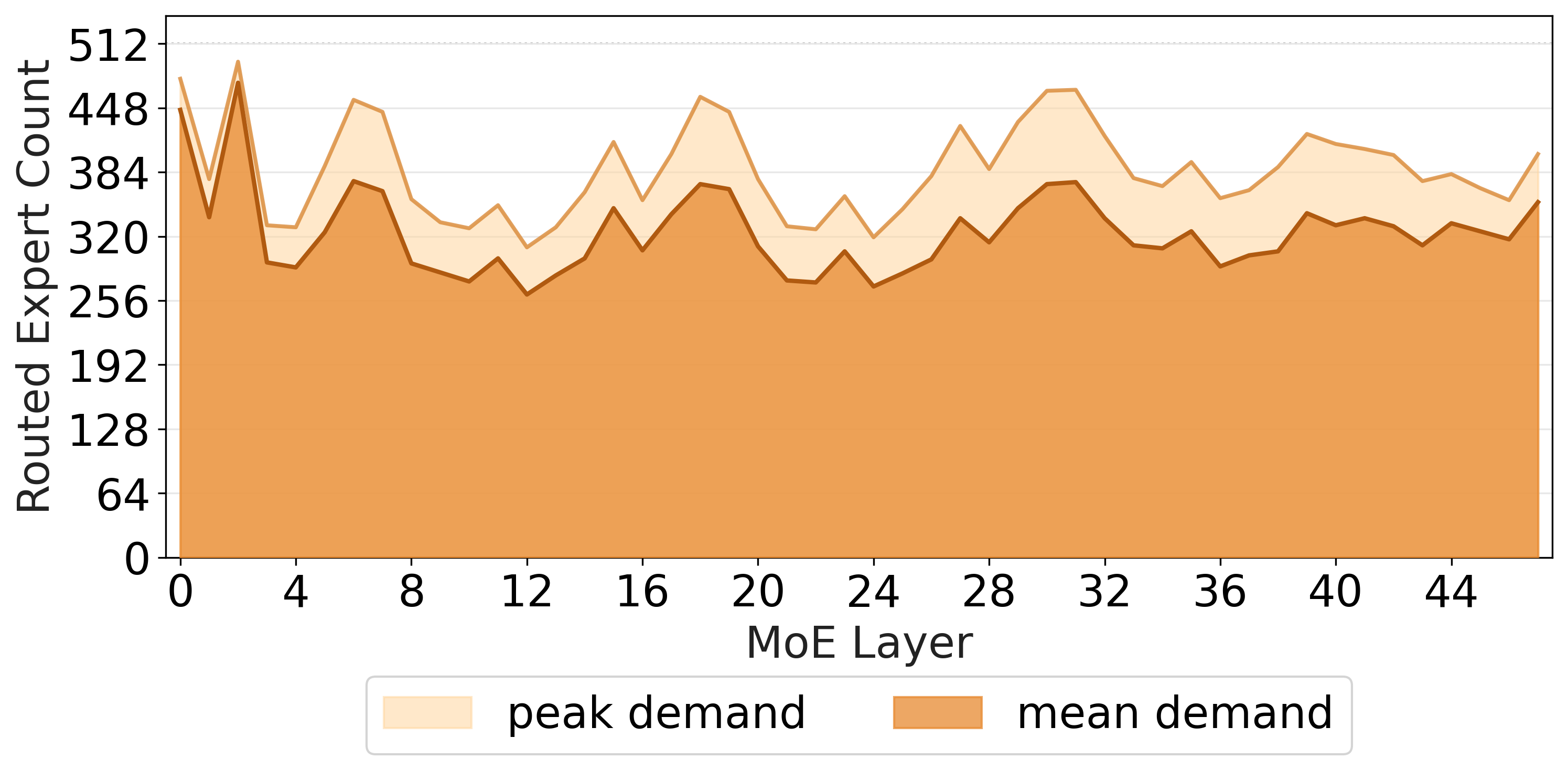}
    \caption{\textbf{Layer-level routed-expert demand under batched
    execution} (Qwen3-Next-80B-A3B-Instruct, TP=4, chunked prefill,
    \texttt{max\_num\_seqs}${=}64$). Across 1{,}012 model steps, the
    per-layer union averages 262--473 of 512 routed experts and reaches
    494, despite top-$k{=}10$ routing per token.}
    \Description{Bar chart with MoE layer index on the x-axis (spanning all 48 layers) and the number of distinct routed experts demanded per layer on the y-axis. Each bar represents one layer. The bar heights vary considerably across layers, with some layers requiring only a modest number of experts and others requiring a much larger union of active experts due to diverse routing across the batch. The chart shows that neither the size nor the composition of the per-layer active expert set is fixed or predictable, motivating the use of deterministic staging rather than on-demand per-expert fetching.}
    \label{fig:layer-demand}
\end{figure}%
}

\newcommand{\figHidingCopy}{%
\begin{figure}[t]
  \centering
  \includegraphics[width=\columnwidth]{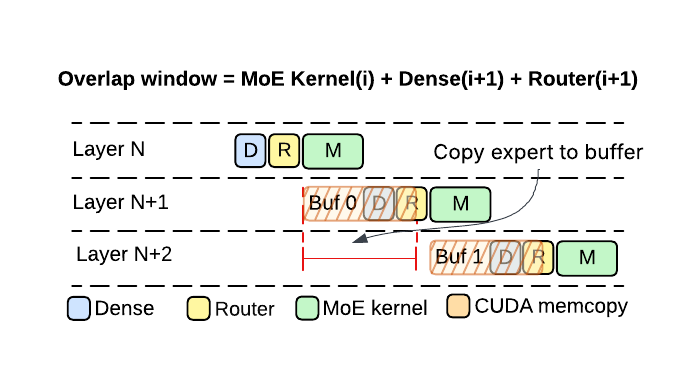}
  \caption{Copy--compute overlap across three consecutive layers.
  The transfer for layer $N{+}1$ begins during layer $N$'s MoE kernel
  and overlaps the computation preceding layer $N{+}1$'s MoE kernel;
  any remaining transfer delays that kernel.}
  \Description{Timeline diagram showing two parallel CUDA streams across three consecutive MoE layers (N, N+1, N+2). The top row (main compute stream) shows sequential operations for each layer: MoE Kernel, Dense, and Router operations drawn as labeled rectangles. The bottom row (copy stream) shows host-to-device PCIe transfers for the next layer's offloaded expert weights, drawn as hatched rectangles that are temporally offset to begin at the start of the current layer's MoE Kernel. A bracket labels the overlap window as approximately 8.8 milliseconds, spanning the MoE Kernel of layer N plus the Dense and Router operations of layer N+1; any part of the transfer not covered by this window delays the next layer's MoE Kernel.}
  \label{fig:hiding_copy}
\end{figure}%
}

\newcommand{\figRatioTradeoff}{%
\begin{figure*}[t]
  \centering
  \includegraphics[width=\textwidth]{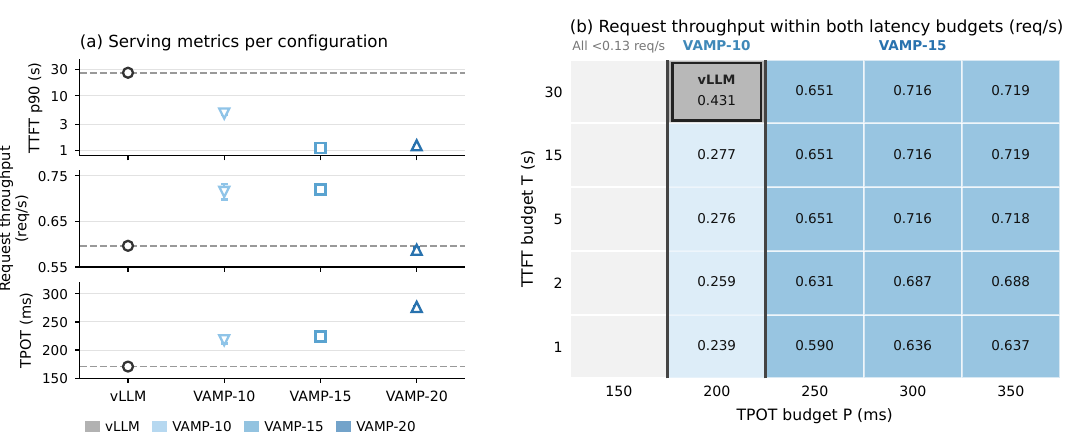}
  \caption{Expert-offloading-ratio tradeoffs on the recorded trace
  ($n{=}5$). Panel (a) reports TTFT p90 (log scale), request
  throughput, and TPOT. Panel (b) compares the configurations over pairs of latency
  budgets ($\mathrm{TTFT}\le T$, $\mathrm{TPOT}\le P$). Outside the
  gray $P{=}150$\,ms column, the band label names the configuration
  with the highest mean request rate satisfying both budgets, and each
  cell reports that rate in req/s.
  The outlined cell marks \vllm{}'s only win
  outside this low-throughput region; every configuration in the gray
  column remains below 0.13\,req/s. At $T\ge15$\,s and $P{=}350$\,ms the winner and
  runner-up overlap within one SD.}
  \Description{Two panels. Panel (a): three stacked mini-axes over
  four configurations. TTFT p90 on a log axis falls from about 26
  seconds for vLLM to 4.7 for VAMP-10 and about 1.1 to 1.2 seconds for
  VAMP-15 and VAMP-20; request throughput rises from 0.596 to 0.714 and
  0.720 for VAMP-10 and VAMP-15 and returns to 0.588 for VAMP-20;
  server-side mean TPOT rises monotonically from 170.9 ms to 217.5,
  224.0, and 276.5 ms. Markers show means with plus-or-minus one
  standard deviation error bars, which are smaller than the markers in
  most cases, and dashed lines mark the vLLM reference. Panel (b): a
  five-by-five grid over TTFT budgets of 1 to 30 seconds and TPOT
  budgets of 150 to 350 ms. Each cell reports the highest mean request rate satisfying both
  latency budgets; a band label above the columns names the
  winning configuration for that region. The 150 ms column is gray,
  with no winner emphasized. VAMP-10 wins the 200 ms column (0.239 to
  0.277) except at 30 seconds, where an outlined cell marks vLLM's only
  win (0.431). VAMP-15 wins every cell at 250 ms and above (0.590 to
  0.719).}
  \label{fig:ratio-tradeoff}
\end{figure*}%
}

\newcommand{\figWTwoSweep}{%
\begin{figure}[t]
  \centering
  \includegraphics[width=\columnwidth]{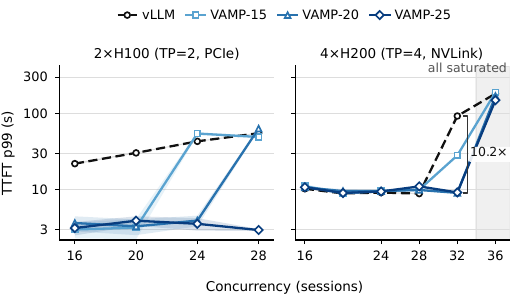}
  \caption{KV-cache capacity boundaries on \hone{} and \htwo{}
  ($n{=}5$, mean$\pm$SD). A larger expert-offloading ratio shifts the
  TTFT cliff to higher concurrency; an insufficient ratio crosses its
  own boundary and can trail \vllm{}.}
  \Description{Two side-by-side line charts of steady-state TTFT p99 in
  seconds versus session concurrency, both on identical log axes from 3
  to 300 seconds. Left panel, two H100 GPUs with tensor parallelism 2,
  spans concurrency 16 to 28: the black dashed vLLM baseline rises
  steadily from 22 to 55 seconds, while VAMP-15, VAMP-20, and VAMP-25
  (light to dark blue) stay near 3 seconds until each ratio's boundary;
  VAMP-15 jumps above the baseline at concurrency 24 and VAMP-20 at 28,
  each marked by a small label reading VAMP-15 limit and VAMP-20 limit,
  while VAMP-25 remains near 3 seconds throughout. Right panel, four H200
  GPUs with tensor parallelism 4, spans concurrency 16 to 36: all four
  series overlap near 9 to 11 seconds through concurrency 28; at
  concurrency 32 the baseline jumps to 93 seconds while VAMP-20 and
  VAMP-25 stay near 9 seconds, spanned by a thin bracket labeled 10.2
  times lower; at concurrency 36, shaded and labeled all saturated, every
  series exceeds 149 seconds.}
  \label{fig:w2-sweep}
\end{figure}%
}

\newcommand{\figDockerRegime}{%
\begin{figure}[t]
  \centering
  \includegraphics[width=\columnwidth]{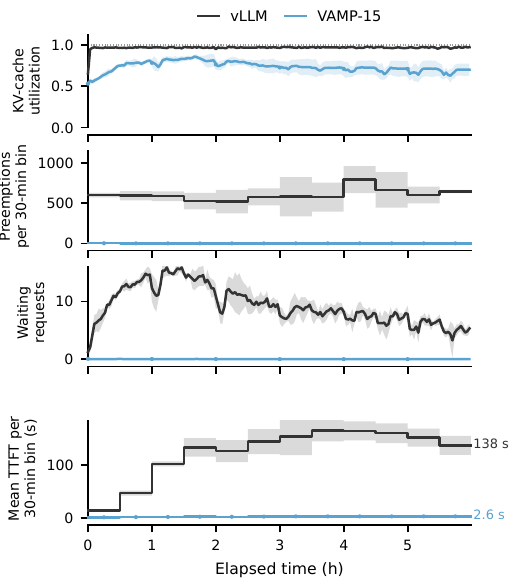}
  \caption{Serving-regime contrast over the 6-hour closed-loop
  SWE-agent execution (mean$\pm$SD over $n{=}3$ runs per system). The
  top three panels show system state---KV-cache
  utilization, preemptions per 30-minute bin, and waiting
  requests---and the bottom panel is the user-visible outcome, mean
  TTFT per 30-minute bin.
  Because closed-loop trajectories differ across systems and runs,
  the comparison characterizes observed serving regimes under the
  same workload setup, not a matched-request effect.}
  \Description{Four stacked time-series panels over zero to six hours
  of elapsed serving time comparing vLLM, in dark gray, and VAMP-15,
  in light blue. Panel one, KV-cache utilization: the vLLM line rises
  immediately to about 0.97 and stays pinned near the capacity line
  for the full six hours, while the VAMP-15 line rises to about 0.8
  and drifts between 0.6 and 0.8 with a visible standard-deviation
  band. Panel two, preemptions per 30-minute bin, drawn as steps: the
  vLLM steps stay between about 500 and 800 for the entire run with
  one taller step near 4.5 hours, while the VAMP-15 steps contain
  only 1--2 warm-up events in the first bin and are zero thereafter.
  Panel three, waiting requests: the vLLM line climbs to
  about 15 in the first 90 minutes and declines slowly toward 5,
  while the VAMP-15 line lies on zero. Panel four, mean TTFT per
  30-minute bin, visually separated below as the outcome panel: the
  vLLM steps climb from about 13 seconds in the first bin to about
  160 seconds by mid-run, ending near 138 seconds, while the VAMP-15
  line stays flat near 2.6 seconds; both final values are labeled at
  the right edge.}
  \label{fig:docker-regime}
\end{figure}%
}

\newcommand{\figDockerThreeArm}{%
\begin{figure}[t]
  \centering
  \includegraphics[width=\columnwidth]{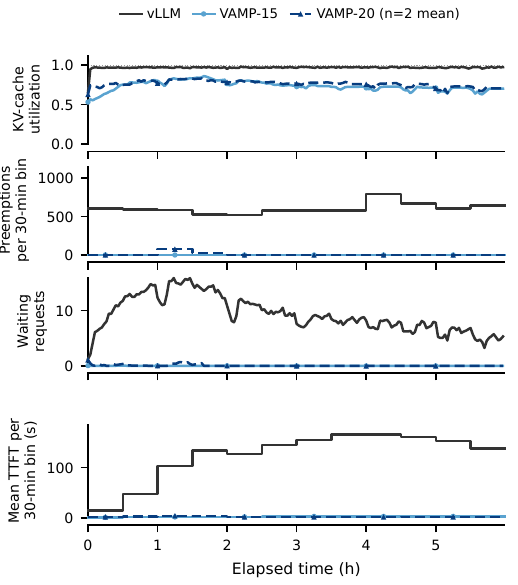}
  \caption{Three-arm view of the closed-loop execution (same panels
  as Figure~\ref{fig:docker-regime}, mean lines only). \vllm{} and
  \sys-15 average $n{=}3$ runs; \sys-20 averages its two
  protocol-identical runs ($n{=}2$)---a third run with a different
  LM-query timeout is excluded as a protocol variant. \sys-15 and
  \sys-20 maintain similar mean KV utilization, so \sys-20 provides
  little additional average KV headroom over the selected \sys-15
  operating point. One of the two \sys-20 runs exhibited a transient
  saturation with a 211-preemption burst around 1--2\,h; the $n{=}2$
  mean dilutes it to the small step visible in the second panel. With
  two runs we report these as observations, not a reproducibility
  claim.}
  \Description{Four stacked time-series panels identical in structure
  to the main closed-loop figure, with a third series added: VAMP-20
  as a dark-blue dashed line with triangular markers roughly every
  hour, alongside vLLM in dark gray and VAMP-15 in light blue with
  circular markers. In the KV-cache utilization panel the VAMP-15 and
  VAMP-20 lines overlap around 0.7 for most of the run while vLLM
  stays pinned near 1.0. In the preemptions panel the vLLM steps stay
  between 500 and 800 while VAMP-15 lies on zero apart from 1--2
  warm-up events in the first bin, and VAMP-20 shows a
  single low step of about 100 between one and two hours, the diluted
  trace of one run's burst. In the waiting-requests panel vLLM climbs
  to 15 and declines while both VAMP lines lie on zero with a brief
  small bump for VAMP-20 near 1.5 hours. In the TTFT panel the vLLM
  steps climb above 100 seconds while both VAMP lines stay flat below
  about 5 seconds.}
  \label{fig:docker-three-arm}
\end{figure}%
}

\newcommand{\figTurnwiseTTFT}{%
\begin{figure}[t]
  \centering
  \includegraphics[width=\columnwidth]{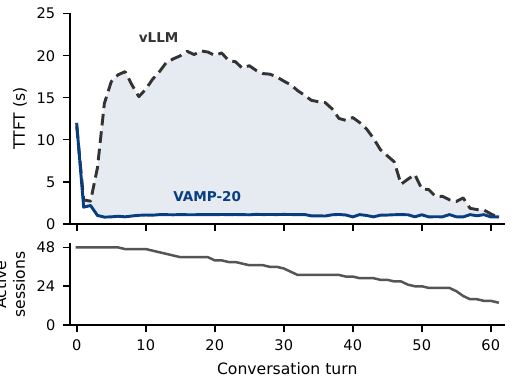}
  \caption{Per-turn median TTFT over the 48-session recorded replay
  (pooled over $n{=}5$ runs per system), shown for \sys-20 from the
  same recorded trace as Figure~\ref{fig:ratio-tradeoff}; the bottom
  panel shows the number of still-active sessions. Turns beyond 61,
  where fewer than 12 sessions remain, are omitted: that
  finite-replay tail is not a steady-state serving regime.}
  \Description{Two stacked panels sharing a conversation-turn axis
  from 0 to 61. Top panel: per-turn median TTFT in seconds on a
  linear axis from 0 to 25. The dark-gray dashed vLLM line and the
  dark-blue solid VAMP-20 line both start near 12 seconds at turn 0;
  the vLLM line then climbs to a plateau around 19 to 20 seconds
  through roughly turn 25 and declines gradually afterward, while the
  VAMP-20 line drops to about 1 second by turn 3 and stays flat; the
  area between the lines is filled in light blue. The two series are
  identified by direct color-matched name labels.
  Bottom panel: the number of active sessions declines from 48 to
  about 14 across the same turns.}
  \label{fig:turnwise-ttft}
\end{figure}%
}

\newcommand{\figDockerCollapse}{%
\begin{figure}[t]
  \centering
  \includegraphics[width=\columnwidth]{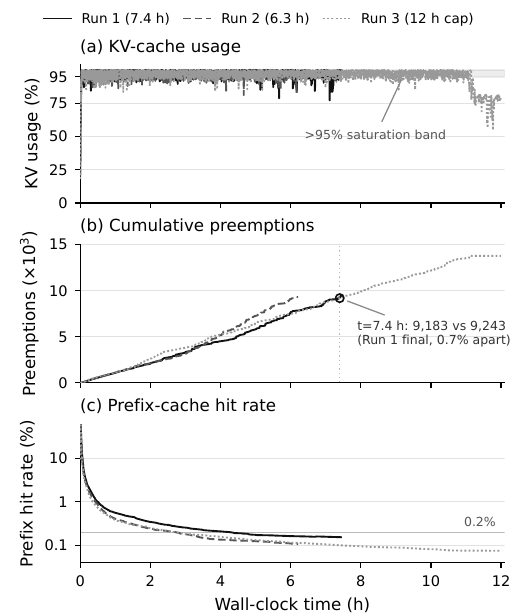}
  \caption{The same collapse signature across three independent baseline
  runs: KV-cache usage pinned above 95\%,
  near-linear preemption accumulation (9,183 vs.\ 9,243 at $t{=}7.4$\,h,
  0.7\% apart), and prefix-cache hit rate decaying below 0.2\%.}
  \Description{Three vertically stacked time-series panels sharing a
  wall-clock x axis from 0 to 12 hours, showing three baseline runs from
  Runs 1, 2, and 3 as black solid, dark gray dashed, and
  light gray dotted lines. Panel (a): KV-cache usage rises immediately to
  a shaded saturation band above 95 percent and stays there for each
  run's full duration; the 12-hour run declines to about 78 percent
  only in its final hour. Panel (b): cumulative preemptions grow
  near-linearly and overlap closely across runs, reaching about 9,000 by
  7.4 hours, annotated where the 12-hour run's count at 7.4 hours, 9,183,
  nearly matches Run 1's final count of 9,243; Run 3
  continues to 13,767 by 12 hours. Panel (c): cumulative prefix-cache hit
  rate on a log scale decays from over 10 percent at the start to below a
  0.2 percent reference line in all three runs.}
  \label{fig:docker-collapse}
\end{figure}%
}

\newcommand{\figLMCacheWindows}{%
\begin{figure}[t]
  \centering
  \includegraphics[width=\columnwidth]{figures/fig_lmcache_bars.pdf}
  \caption{Latency--completion comparison on the LMCache trace
  ($n{=}3$; bars = mean, error bars = SD, dots = individual runs).
  (a)~Steady-state TTFT p99; (b)~share of recorded turns completed
  within the 60-minute window, at 64 and 128 sessions.}
  \Description{Two grouped-bar panels comparing vLLM in dark gray,
  VAMP-15 in light blue, and VAMP-20 in dark blue, grouped by 64 and
  128 sessions with a shared legend on top; every bar carries its
  mean value, an SD error bar, and three per-run dots. Panel (a),
  steady-state TTFT p99 on a linear axis from 0 to about 33 seconds:
  at 64 sessions the bars read 22, 1.6, and 1.3 seconds; at 128
  sessions they read 29, 4.0, and 4.3 seconds. Panel (b), turns
  completed within 60 minutes on a 0 to 100 percent axis: at 64
  sessions the bars read 99, 97, and 96 percent; at 128 sessions they
  read 95, 72, and 67 percent.}
  \label{fig:lmcache-windows}
\end{figure}%
}

\newcommand{\figReplayProgress}{%
\begin{figure}[t]
  \centering
  \includegraphics[width=\columnwidth]{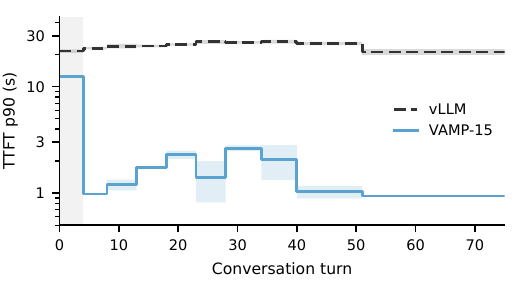}
  \caption{Progress-aligned TTFT p90 over ten equal-observation bins,
  each aggregating approximately 210 matched (task,\,turn) pairs
  ($n{=}5$; band $\pm$1\,SD). Bins are plotted as steps over their
  actual recorded-turn ranges, so unequal bin widths are visible; the
  shaded first bin ($[0,4)$) includes the synchronized turn-0
  requests from all 48 sessions.}
  \Description{Step chart of TTFT p90 in seconds on a log axis over a
  conversation-turn axis from 0 to 75, with steps spanning each bin's
  actual turn range; the final step, spanning $[51,75)$, is
  visibly wider than the others. The dark-gray dashed vLLM steps stay
  between 21 and 26 seconds across the whole axis. The light-blue
  VAMP-15 steps start at 12.4 seconds in the lightly shaded first bin
  spanning the interval $[0,4)$, which includes the synchronized
  turn-0 requests of all 48 sessions, then stay between 0.9 and 2.6 seconds
  for the rest of the axis, roughly an order of magnitude below vLLM.
  Step-shaped bands show one standard deviation across five runs.}
  \label{fig:replay-progress}
\end{figure}%
}

\newcommand{\figQSurfaceFacets}{%
\begin{figure*}[t]
  \centering
  \includegraphics[width=\textwidth]{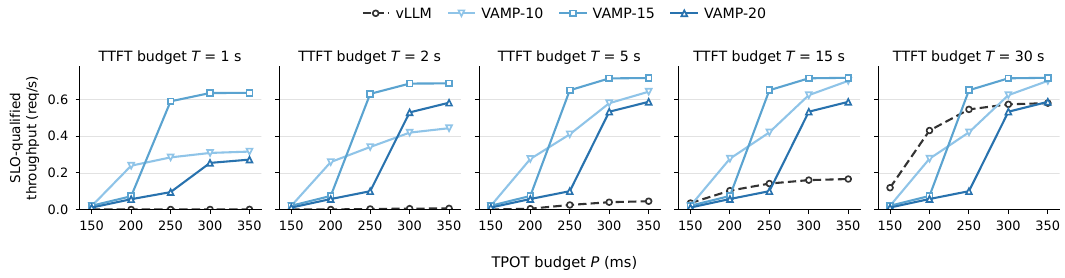}
  \caption{Underlying SLO-qualified-throughput slices for the evaluated
  $5\times5$ $Q(T,P)$ grid of Figure~\ref{fig:ratio-tradeoff}(b): each
  panel fixes the TTFT budget $T$ and sweeps the TPOT budget $P$ for
  all four configurations ($n{=}5$ run means).}
  \Description{Five side-by-side line charts, one per TTFT budget of 1,
  2, 5, 15, and 30 seconds, each plotting SLO-qualified throughput in
  requests per second against TPOT budgets from 150 to 350 milliseconds for
  four configurations. In the first four panels the black dashed vLLM
  line stays near zero or below 0.17 while VAMP-15 rises steeply at 250
  milliseconds to about 0.6 to 0.72; VAMP-10 rises earlier but plateaus
  lower, and VAMP-20 rises only at 300 milliseconds. In the 30-second
  panel the vLLM line reaches 0.43 at 200 milliseconds, exceeding the
  VAMP configurations there, before VAMP-15 overtakes it from 250
  milliseconds onward.}
  \label{fig:qsurface-facets}
\end{figure*}%
}

\newcommand{\figHBMBudget}{%
\input{fig-intro-hbm}%
}

\begin{abstract}
Long-running multi-turn requests accumulate reusable key-value (KV) state. Once this state exceeds a fixed GPU KV-cache allocation, serving systems evict reusable prefixes, repeat prefill work, and may preempt requests. This pressure is particularly acute for Mixture-of-Experts (MoE) models: their expert weights occupy most GPU high-bandwidth memory (HBM), even though each token activates only a sparse subset of experts. A static boundary between weights and the KV cache prevents serving systems from using expert memory to preserve reusable state as conversations grow.

We present \sys, an MoE serving framework that changes this HBM boundary at runtime. When a KV allocation cannot be satisfied, \sys compares the estimated future work of three alternatives: staging expert weights from host memory, evicting cached prefixes that may require re-prefill, or preempting and rescheduling requests. It then converts a bounded expert-weight region into KV-cache capacity when that action has the lowest estimated penalty. CUDA Virtual Memory Management page remapping performs this conversion without copying resident KV data.

We implement \sys in the \vllm serving engine and evaluate Qwen3-Next-80B on recorded and controlled multi-turn workloads. Across five replays of a recorded 2{,}103-turn SWE-bench agent workload, \sys with a 15\% maximum expert-offloading ratio reduces time-to-first-token (TTFT) p90 from 26.1\,s to 1.10\,s (23.6$\times$) and increases request throughput by 20.7\% relative to unmodified \vllm{}, while increasing time per output token (TPOT) by 31.1\%.
\end{abstract}

\section{Introduction}
\label{sec:intro}

Large language models (LLMs) are now deployed at scale across a growing range of applications, from code generation to multi-turn agentic tasks, making efficient GPU memory management a central challenge in production serving. Modern LLM serving systems process requests by batching multiple inputs together to sustain high throughput, while retaining key-value (KV) caches to avoid recomputing previously processed tokens. Both strategies impose heavy pressure on GPU high-bandwidth memory (HBM): model weights must remain resident while the KV-cache footprint grows linearly with batch size and context length, making GPU memory capacity the primary bottleneck in practical serving deployments.

\figThesis

Existing systems typically treat model placement and KV-cache management as separate decisions. For instance, \vllm~\cite{kwon2023vllm} configures model-weight placement at engine initialization, while allocating KV-cache blocks at runtime from a preallocated KV-cache pool.
Implicit in this separation are two assumptions: (1) the model-weight footprint occupies only a modest fraction 
of HBM, leaving ample 
room for KV cache, and (2) the KV-cache requirements remain strictly within the limits of the preallocated pool.
These assumptions generally hold in dense-model serving scenarios with short context lengths, and static memory partitioning is a sufficient strategy.

However, recent developments undermine both assumptions. Mixture-of-Experts (MoE) architectures invalidate the first assumption by structurally decoupling model capacity from per-token compute cost: because only a sparse subset of experts activates per token, adding more experts increases total parameter count without increasing inference FLOPs~\cite{fedus2022switch}. This creates a strong incentive to scale total model size aggressively -- quality improves with more experts while inference compute stays nearly constant~\cite{abnar2025paramsvflops}. The result is an explosive growth in total parameter counts across successive MoE generations: Mixtral-8x7B reaches 47B total parameters with only ${\sim}$13B active per token~\cite{kamahori2025fiddler}; DeepSeek-V3 scales to 671B total with 37B active, an 18$\times$ ratio~\cite{deepseekv3}; our target model Qwen3-Next-80B carries 80B total parameters while activating only 3B per token~\cite{qwen2025qwen3next}.
Unlike dense models, where weight memory scales modestly with quality, MoE weight memory now dominates HBM and will continue to grow as models scale.

The second assumption is invalidated in
multi-turn agentic settings, 
where the KV-cache footprint expands continuously as agents accumulate context over successive turns, often keeping KV-cache utilization near saturation (80\% to 100\%) and causing repeated evictions and re-prefills that collapse cache efficiency well before memory is exhausted~\cite{chen2026concur}.
Production traces confirm the scale of this growth: agentic trajectories average 157 turns and 32.7k tokens of context, with some runs exceeding 1M tokens~\cite{wu2026dualpath}.
Ultimately, when MoE models are used for multi-turn agentic tasks, both assumptions no longer hold. 

Under sustained memory pressure, KV-cache pool overflow triggers a cascade of evictions and request preemptions, forcing repeated recomputation and rescheduling that further compound memory pressure. Such a collapse is not gradual, but 
abrupt and threshold-driven. As illustrated in Figure~\ref{fig:thesis}(b), once the reusable prefix-cache footprint outgrows the KV-cache capacity, the prefix-cache hit rate collapses and the 99th-percentile (p99) time-to-first-token (TTFT) rises sharply. 
We call this phenomenon the \textbf{prefix-cache cliff}. Static HBM partitioning causes the cliff to occur earlier because memory assigned to expert weights cannot be reallocated to the KV cache.
Moving the prefix-cache cliff to higher KV-cache demand extends the operating range over which MoE serving can sustain low latency.

This shift requires dynamically reallocating HBM from the expert-weight region to the KV cache.
In MoE serving, expert weights consume a substantial portion of HBM, even though
any single token activates only a small fraction of them.
Offloaded expert weights can instead be kept in host memory and staged before execution regardless of which experts are activated, freeing the corresponding HBM pages for KV-cache allocation.

Yet, enabling such cross-pool reallocation at runtime presents two fundamental challenges.
\textbf{First}, the future cost of each allocation choice is unknown at decision time. Expert offloading incurs repeated weight transfers during subsequent execution, so its total cost depends on how much work remains while the weights stay offloaded.
Evicting a KV-cache block incurs a re-prefill penalty, whose magnitude is contingent upon whether future requests would have benefited from its reuse -- a factor determined by workload-level prefix sharing~\cite{zheng2024sglang}.
Without forward-looking estimates, the system cannot distinguish a low-cost eviction from a prohibitively expensive one.
\textbf{Second}, existing serving frameworks enforce a static memory partition between model weights and the KV-cache pool, lacking the mechanisms for dynamic boundary adjustment at runtime.
Under memory pressure, the current schedulers are restricted to either evicting cached KV blocks or preempting running requests~\cite{kwon2023vllm}, even when expert offloading would be less costly than either alternative. While prior offloading strategies~\cite{kamahori2025fiddler, cao2025moelightning} migrate experts to host CPU memory to accommodate large models, they do not repurpose the reclaimed HBM for KV cache or dynamically coordinate between these two memory pools.

To address these challenges, we propose Value-Aware Memory Partitioning (\sys), an LLM serving framework that mitigates the prefix-cache cliff by unifying GPU memory (which is HBM in our setting) across expert weights and the KV cache with dynamic management. Instead of static memory partitioning, \sys employs a dynamic allocation policy guided by a \emph{three-way cost model}. This cost model compares the estimated cost of each action at allocation time: weight transfer for expert offloading, re-prefill for KV-cache eviction, and rescheduling for request preemption. In order to support the \sys design without incurring data copying overhead during repartitioning, we employ CUDA Virtual Memory Management (VMM)~\cite{CUDA-VMM} to remap physical HBM pages at runtime between the expert-weight region and the KV-cache region at page granularity, enabling fine-grained cross-pool reallocation.
Furthermore, \sys overlaps the transfers of offloaded expert weights with computation through double buffering, reducing the exposed transfer latency.
In particular, this paper makes the following contributions:
\squishlist
    \item \textbf{A Three-Way Cost Model:} We develop an analytical model that estimates the penalties for (i) expert offloading, (ii) KV-cache eviction, and (iii) request preemption. By integrating %
    prefix-cache hit rates and block recency, the model selects the plan with the lowest estimated cost for each allocation.
    \item \textbf{The \sys Framework:} We propose \sys, a serving framework that dissolves the static boundaries between model weights and the KV cache. \sys manages GPU HBM as a shared resource, reallocating pages from the expert-weight region to the KV-cache pool as demand grows. On our H100 testbed, \sys expands the per-GPU KV-cache pool by up to 2.7$\times$.
    \item \textbf{Comprehensive Experimental Analysis:} We integrate \sys into the \vllm serving engine and comprehensively analyze the performance of \sys compared with \vllm on Qwen3-Next-80B under various multi-turn workloads. On a 2{,}103-turn recorded-trace replay, \sys reduces TTFT p90 from 26.1\,s to 1.10\,s (23.6$\times$) and increases request throughput by 20.7\% relative to unmodified \vllm{}, while increasing time per output token (TPOT) by 31.1\%.
\squishend

\section{Background}
\label{sec:background}

\subsection{KV Cache}
\label{subsec:kv_bg}

LLMs generate tokens autoregressively, computing attention over all prior tokens at each step.
To avoid recomputing key and value projections, they are cached in GPU memory as the \emph{KV cache}, which grows linearly with context length and batch size: a single 10K-token request on Qwen3-8B occupies roughly 1.4\,GiB; at typical serving concurrency with multi-turn workloads (e.g., SWE-bench~\cite{jimenez2024swebench}), total KV-cache footprint can easily reach tens of GiB.

\textbf{Prefix caching} retains KV cache after a request completes so future requests sharing a common prefix (e.g., a system prompt) can skip recomputation.
Under memory pressure, cached blocks are evicted; any evicted block later needed triggers a \emph{re-prefill}, reprocessing the corresponding tokens from scratch at a significant latency cost.
Figure~\ref{fig:kv_cache_dialog} illustrates this dynamic across a
multi-turn agentic session. Because each turn extends the previous
context, the blocks cached during turns 1--3 are reused verbatim by
every subsequent turn, so the prefill cost of a turn is limited to its
newly appended tokens. The same growth, however, eventually
exhausts the KV-cache pool: in turn~4, the allocator evicts the
oldest blocks to admit the new request, and the evicted blocks are
needed again by the next turn. The result is a cache miss over the
shared prefix, a re-prefill of tokens that were already processed, and
a TTFT spike---the eviction decision is made without regard to how
likely, or how expensive, the evicted blocks are to be reused.

\subsection{Mixture-of-Experts Architecture}
\label{subsec:moe_bg}

Mixture-of-Experts (MoE) models replace the dense feed-forward sublayer in each Transformer block with $N$ expert feed-forward networks (FFNs) and a learned router.
For each token, the router selects the top-$k$ experts (with $k=10$ in our setting) to process that token; the remaining $N-k$ experts are not activated.
This \emph{conditional computation} enables total parameter counts to scale to hundreds of billions while keeping the per-token FLOPs comparable to a much smaller dense model.

Although each token activates only $k$ experts, the model contains weights for all $N$ experts, and different tokens may select different subsets. Consequently, the total expert-weight footprint can greatly exceed the per-token active footprint, placing substantial pressure on GPU HBM.

\figMotivation

\subsection{Memory Offloading}
\label{subsec:offload_bg}

When GPU memory is insufficient to hold all required data, systems offload objects to slower memory tiers, typically CPU DRAM.

\textbf{Weight offloading} moves model parameters (e.g., expert FFN weights) from GPU to CPU. When an offloaded weight is needed, it must be transferred back before the corresponding computation can proceed. The transfer latency is bounded and one-time per fetch, and can be hidden by overlapping transfers with computation on resident weights.

\textbf{KV-cache offloading} moves KV-cache blocks from GPU HBM to CPU DRAM or storage, preserving them for reuse across requests~\cite{liu2025lmcache,hu2024memserve,qin2024mooncake}. When the offloaded blocks are needed again, the system restores them to GPU memory instead of recomputing the corresponding tokens; both offloading and restoring the blocks incur transfer overhead.

\section{Memory Management Challenges}
\label{sec:challenges}

Existing LLM serving systems draw a hard boundary between memory allocated to model weights and the KV-cache pool, managing each independently under a fixed budget. This design relies on two assumptions that generally held for early dense models: loading the model leaves enough HBM for the KV cache, and KV-cache demand stays bounded within the preallocated pool. Neither assumption holds for modern MoE models serving multi-turn agentic workloads. MoE architectures cause expert weights to dominate HBM, while multi-turn agentic workloads drive KV-cache demand far beyond what a statically sized pool can absorb. Together, these trends motivate cross-pool reallocation as KV-cache demand grows. For example, a system can offload a subset of expert weights and remap the corresponding HBM pages into the KV-cache pool. Supporting such reallocation raises two challenges. First, existing serving engines provide no runtime mechanism for reallocating capacity across fixed pool boundaries (\S\ref{sec:rigid}). Second, even with such a mechanism, the future cost of each allocation choice is unknown at decision time, so the system cannot determine which data is most valuable to retain in HBM (\S\ref{sec:blind}).

\subsection{Assumptions Invalidated by MoE Workloads}

\noindent \textbf{Assumption 1: A substantial fraction of HBM is available for the KV cache.}
Dense models validated this assumption in practice. A single H100 provides 94\,GiB of HBM. Llama-3.1-8B occupies about 15\,GiB in BF16, roughly 16\% of the device. Even Llama-3.1-70B, spread across 2$\times$H100 under tensor parallelism (TP2), occupies about 130\,GiB of the 188\,GiB aggregate, roughly 69\%. In both cases, a substantial fraction of HBM remains available beyond the model weights.

Mixture-of-Experts architectures break this property structurally. MoE models scale capacity by maintaining many expert feed-forward networks and routing each token to a small subset at inference time, as illustrated in Figure~\ref{fig:motivation}(a). The result is a large and growing gap between total parameter count and active parameter count. Mixtral-8x7B has 47B total parameters while activating only ${\sim}$13B per token, a 3.6$\times$ disparity~\cite{jiang2024mixtral}. DeepSeek-V3 widens this to 671B total versus 37B active, an 18$\times$ ratio~\cite{deepseekv3}. On our target model, Qwen3-Next-80B holds 80B parameters while activating only 3B per token~\cite{qwen2025qwen3next}.

The consequence is that weight memory now dominates HBM. Serving Qwen3-Next-80B on 2$\times$H100 under TP2, expert weights alone consume 72.0\,GiB of the 94\,GiB available per GPU; after accounting for the remaining model state, runtime overhead, and reserved headroom, the baseline KV-cache pool is only 8.02\,GiB per GPU (Table~\ref{tab:platforms}; Figure~\ref{fig:thesis}(a)).
MoE scaling increases the total expert count while keeping the per-token active count roughly constant~\cite{deepseekv3, qwen2025qwen3next}, so expert weights occupy a growing share of HBM as models scale.

\noindent \textbf{Assumption 2: KV-cache requirements remain within a preallocated pool.}
This assumption was usually sufficient for the single-turn request patterns targeted by existing serving systems. Each request carried an independent context of a few hundred to a few thousand tokens; the KV-cache pool was sized once at initialization, and demand rarely exceeded it. When eviction was necessary, displacing one or two least-recently-used (LRU) blocks restored headroom without materially affecting throughput.

Agentic workloads violate this assumption structurally. In multi-turn settings such as code assistants, tool-use agents, and long-horizon planners, the session context is cumulative: each turn appends the prior turn's output, tool results, and observations before the next prefill, so both KV-cache footprint and reusable prefix length grow across turns, as shown in Figure~\ref{fig:motivation}(b). 

Under sustained pressure, this behavior becomes self-reinforcing: as
the prefix-cache hit rate falls, large re-prefills inflate TTFT and
consume serving capacity, sustaining memory pressure and further
preemptions (\S\ref{sec:eval-stability}).

Overflow under these conditions is qualitatively different from the occasional eviction that a static design handles gracefully. Once the pool saturates, every new block allocated displaces an existing one. Prefix-cache hit rate collapses as blocks evicted from one turn are immediately needed by the next. This is not a smooth performance degradation; it is a threshold effect. As Figure~\ref{fig:thesis}(b) illustrates, TTFT remains stable until the pool saturates and then rises sharply.
We call this the \emph{prefix-cache cliff}.

\paragraph{Both assumptions fail for MoE models serving multi-turn agentic workloads}
MoE models serving multi-turn agentic workloads violate both assumptions at once. Expert weights claim most of HBM, leaving little room for KV cache, while growing session context continuously increases demand until the pool overflows. The static partition provides no escape: the scheduler cannot borrow weight memory to expand KV capacity, and its only available tools, LRU KV-cache eviction and request preemption, reduce prefix-cache coverage precisely when prefix reuse is highest, triggering an eviction--recompute cascade.

\subsection{Issues with Static Memory Partitioning}
\label{sec:rigid}

Existing serving engines partition HBM into two disjoint regions at startup: a weight pool sized to hold the full model, and a KV-cache pool sized from whatever HBM remains. Neither pool can grow at the expense of the other at runtime.

In \vllm~\cite{kwon2023vllm}, the KV-cache pool is preallocated as a fixed block at engine initialization. Its size is determined once, based on the HBM remaining after model weights are loaded, and never changes during serving. Under memory pressure, the scheduler operates entirely within this pool: it reclaims space by evicting cached blocks in LRU order, or as a last resort preempts running requests. There is no code path by which the scheduler can reclaim a byte from the weight region and redirect it to KV cache.

This static partitioning is wasteful for MoE models. Although expert activation is sparse at inference time, the serving system still reserves HBM for the full expert weight set. A natural pressure-relief valve is to offload a subset of expert weights to CPU DRAM and immediately repurpose the freed HBM as KV-cache blocks. Prior works such as Fiddler~\cite{kamahori2025fiddler} and MoE-Lightning~\cite{cao2025moelightning} do offload expert weights to CPU DRAM, but neither redirects the freed HBM to the KV-cache pool. When an expert is offloaded, the released memory either remains idle or is reclaimed by the weight region for the next expert to be paged in. The two pools remain disjoint: freeing weight memory does not expand KV-cache capacity.

Oneiros~\cite{li2025oneiros} is the closest prior work in its use of virtual-memory remapping for cross-pool reallocation: it remaps physical GPU memory pages from model parameters to the KV-cache pool using virtual memory primitives. However, Oneiros is designed for multi-tenant dense-model serving: it reclaims memory from \emph{other tenants'} idle model replicas, not from the active model's own expert weights. It does not address MoE architectures, and its remapping mechanism requires a dormant model instance on the same GPU as the source of reclaimable pages. In our setting, there is only one model and there are no idle replicas; the reclaimable memory must come from the serving model's own expert weight pool, a case Oneiros does not handle.

\figKVCacheDialog

The problem intensifies under multi-turn agentic workloads, where KV-cache footprint accumulates steadily across turns~\cite{chen2026concur}. Once the KV-cache pool can no longer satisfy new allocations, the only available actions are cached-block eviction and request preemption. Under static partitioning, neither action can expand KV capacity. Instead, both reduce prefix-cache coverage precisely when prefix reuse is highest. Concretely, the early turns of a session accumulate cached blocks that later turns reuse without recomputation; once the pool is full, admitting a new turn evicts exactly those blocks, so the next turn that extends the same prefix must recompute them and the recomputed blocks consume the space that was just freed. Breaking this cycle requires a shared pool that expands KV-cache capacity with HBM reclaimed from expert weights when expert offloading has the lowest estimated future cost.

\subsection{Issues with Cost-blind Evictions}
\label{sec:blind}

Even with cross-pool reallocation, a second challenge remains: the system must decide \emph{which} data to evict, and each candidate action carries a future penalty whose magnitude is unknown at decision time.

Consider evicting a cached KV block. The penalty is a re-prefill: if a future request shares the evicted prefix, those tokens must be recomputed from scratch at additional TTFT cost proportional to the number of evicted tokens. As shown in Figure~\ref{fig:motivation}(c), when a running request requires a new KV-cache block, the allocator may treat cached blocks as eviction candidates rather than prioritizing free blocks. Whether this re-prefill is triggered depends entirely on whether a future request will match the evicted prefix, a quantity that depends on the workload's prefix-sharing pattern~\cite{zheng2024sglang}. In a multi-turn session where later turns repeatedly access accumulated context, evicting a cached block from the session history means paying re-prefill cost on every subsequent turn that builds on that prefix. In a workload with low prefix reuse, the same eviction costs nothing.

Expert offloading has a different future cost. While expert weights remain offloaded, executing their layers requires repeated host-to-device transfers. The cost of each transfer is bounded, but the total cost depends on how much execution remains. The scheduler must therefore estimate, at allocation time, the span over which this transfer cost will recur.

The consequence is that without forward-looking cost estimates, the system cannot make value-aware eviction decisions. The system has no way to distinguish a cached block that will prevent ten re-prefills from one that will never be reused again, nor an offloading action whose transfer cost recurs over only a few remaining steps from one whose cost persists across many. The inability to estimate future penalty means the system cannot compare the cost of evicting expert weights against the cost of evicting KV-cache blocks, which is a prerequisite for any cross-pool reallocation policy.

\section{\sys{} Design}
\label{sec:design}

\subsection{Overview}
\sys combines three components, shown in
Figure~\ref{fig:overview}: value-aware action selection
(\S\ref{sec:design-decision}), a dynamic resource orchestrator that
repartitions HBM between expert weights and the KV cache
(\S\ref{sec:design-orchestration}), and batching-aware expert
offloading (\S\ref{sec:design-staging}). Rather than preserving a
fixed partition, \sys treats HBM as a shared pool.
When enough KV-cache blocks are free,
\sys allocates them directly. Otherwise it estimates the future
penalties of expert offloading, prefix-cache eviction, and request
preemption, and selects the action with the lowest estimate.

\figOverview

\subsection{Value-Aware Decision}
\label{sec:design-decision}

\sys compares three actions by their estimated future penalties:
(1) evicting prefix-cache blocks, (2) offloading a portion of the MoE
expert weights to host DRAM and reallocating the freed HBM to the KV
cache, and (3) preempting an active request to reclaim KV-cache space.

The relative costs of these actions depend on the serving condition.
When observed prefix reuse is high, expert offloading can preserve
cached prefixes that are likely to be reused, avoiding the re-prefill
work that their eviction would cause. When reuse is low, prefix-cache
eviction is less costly because the evicted prefix blocks are unlikely
to be reused.

\sys estimates each action's future cost from the current cache and
request state, without predicting future expert activations or request
arrivals. We define these costs below.

\paragraph{Cached-block eviction cost ($C_c$)}
The cost of evicting prefix-cached blocks depends on whether future requests would have reused them.
\sys considers cached blocks in LRU order and estimates an \emph{effective reuse probability} using the sliding-window prefix-cache hit rate $r_{\text{hit}}$ over the most recent requests and the age of the LRU-front candidate:
\begin{equation}
p_{\text{eff}} \;=\; \frac{\alpha \cdot r_{\text{hit}}}{1 + \mathit{age}/\tau},
\label{eq:peff}
\end{equation}
where $\alpha$ scales the observed hit rate and $\tau$ controls how quickly the reuse estimate decreases with age.
\sys uses the age of the LRU-front candidate as a proxy for the reclaimed set. Under this model, the remaining candidates have reuse estimates no smaller than the front candidate's, so the approximation tends to underestimate $C_c$ rather than favor cache protection.

Evicting the selected cached blocks discards the KV entries for their tokens. If later requests reuse the corresponding prefixes, those entries must be recomputed during prefill. \sys therefore estimates the eviction cost as
\begin{equation}
C_{c} \;=\; p_{\text{eff}} \cdot \left( n_{\text{tok}} \cdot
t_{\text{prefill}} + t_{\text{overhead}} \right),
\label{eq:cc}
\end{equation}
where $n_{\text{tok}}$ is the number of tokens held by the blocks reclaimed to cover the allocation gap, $t_{\text{prefill}}$ is the configured per-token re-prefill coefficient, and $t_{\text{overhead}} = t_{\text{sched}} + t_{\text{queue}}$ covers scheduling and re-queueing overhead (configured values in \S\ref{sec:Imple}).

\paragraph{Expert-offload cost ($C_e$)}
Expert offloading creates repeated transfer overhead: while
expert weights remain offloaded, each model step stages each layer's
offloaded region before that layer's MoE execution, independently of
the router's selections (\S\ref{sec:design-staging}). Under
continuous batching, a model step may mix prefill and decode tokens.
The cost model uses a decode-based accounting horizon $w$, reflecting
that decode-dominated steps typically offer a shorter computation
window for overlapping these transfers than prefill-heavy steps.

Let $n_{\text{cand}} = g \cdot n_{\text{groups}}$ denote the
shard-equivalent size of the candidate expansion across layers, where
$n_{\text{groups}}$ is estimated from the current allocation gap and
each group contains $g$ shards
(\S\ref{sec:design-orchestration}). \sys estimates the per-step cost
as $\min(n_{\text{cand}}, E) \cdot c_{\text{eff}}$, where $E$ is the
number of local expert shards per layer on each TP rank and
$c_{\text{eff}}$ is a configured coefficient used to compare actions,
not a measured transfer time (\S\ref{sec:Imple}). The cap at $E$
limits only the value used for action comparison; it does not bound
the amount of weight data physically staged.

\sys accounts for this per-step cost over a span of $w$ decode steps:
\begin{equation}
C_{e}(n_{\text{cand}}) \;=\; \min(n_{\text{cand}},\, E) \cdot c_{\text{eff}} \cdot w.
\label{eq:ce}
\end{equation}
Here, $w$ is the smallest remaining decode budget among running
requests, clamped to a configured maximum (\S\ref{sec:Imple}).

\paragraph{Request preemption cost ($C_p$)}
Preemption immediately adds uncached capacity only through the
victim's last partially filled block. Its full blocks enter the cached
queue at its most-recently-used end rather than the uncached pool, and
are considered by $C_c$ only if they later become eviction candidates.

The estimated cost of one preemption is
\begin{equation}
C_{p}^{(1)} =
t_{\text{sched}} + t_{\text{queue}} + n_{\text{partial}} \cdot t_{\text{prefill}},
\label{eq:cp1}
\end{equation}
where $n_{\text{partial}}$ is the number of tokens in the victim's
last partially filled KV-cache block. If the victim's computed tokens
end exactly on a block boundary, preemption provides no uncached block
and the request is not a candidate.

Since one eligible preemption provides one uncached block, \sys
estimates the cost of covering a $G$-block shortfall as
\begin{equation}
C_{p} = C_{p}^{(1)} \cdot G,
\label{eq:cp}
\end{equation}
where $G$ is the number of additional uncached KV-cache blocks
required by the current allocation.
Here, $C_p^{(1)}$ is evaluated for the request that the scheduler
would preempt next under its active scheduling policy. Scaling it by
$G$ uses this next-victim cost as the per-block estimate, putting
preemption on the same $G$-block basis as the other actions.

\paragraph{Combined-action cost ($C_b$)}
Expert expansion proceeds in fixed-size increments. When the required
expansion is not an exact multiple of this increment, rounding it down
leaves part of the allocation gap uncovered. \sys therefore also
considers combining the rounded-down expansion with cached-block
eviction:
\begin{equation}
C_{b} \;=\; C_{e}^{\text{floor}} + C_{c}^{\text{remainder}},
\label{eq:cb}
\end{equation}
where $C_{e}^{\text{floor}}$ is the cost of the rounded-down expansion
and $C_{c}^{\text{remainder}}$ the eviction cost for the remaining gap.

\sys selects the lowest-cost feasible plan for covering the allocation
gap: expert offloading ($C_e$), cached-block eviction ($C_c$), request
preemption ($C_p$), or expansion combined with eviction ($C_b$). If
expert offloading has reached its limit, only $C_c$ and $C_p$ remain.

The controller evaluates these plans from the current cache and
scheduler state using the configured coefficients in
\S\ref{sec:Imple}; $p_{\text{eff}}$ adjusts the eviction-cost estimate
according to observed prefix reuse. In
\S\ref{sec:eval-decisions}, we trace the resulting cost estimates and
decisions below and at the KV-cache capacity boundary.

\subsection{Dynamic HBM Repartitioning}
\label{sec:design-orchestration}
\sys dynamically repartitions physical HBM pages between the expert-weight region and the KV-cache pool.
For each KV-cache allocation, the scheduler computes the number of additional uncached KV-cache blocks it requires and translates this gap into a target number of \emph{expert groups} according to the system's page geometry.
An expert group is the minimum page-aligned remap unit. Its granularity is derived at initialization from the per-rank shard size and the CUDA VMM allocation granularity (\S\ref{sec:Imple}). Grouping determines only which expert shards are remapped together; it does not change the experts selected by the router.

When the cost model selects expert offloading to cover the allocation gap, \sys uses CUDA VMM to unmap the physical pages backing the selected expert groups and remap them into the KV-cache pool.
The expert tensors' virtual address ranges remain reserved, so resident weights retain their original addresses while only the physical backing of the offloaded region is reassigned.

\paragraph{Offloading bound}
On each TP rank and offload-eligible layer, \sys offloads at most $E_{\text{buffer}}$ expert shards and keeps at least $E - E_{\text{buffer}}$ resident, where $E_{\text{buffer}}$ is the number of rank-local expert shards the staging buffer (\S\ref{sec:design-staging}) can hold.
The configured maximum offloading ratio \sys-$r$ (\S\ref{sec:eval-setup}) can tighten this limit further.
Within each offload-eligible layer, \sys keeps a low-index region of expert groups resident and offloads a contiguous high-index region. In the evaluated configuration, it moves the residency boundary uniformly across all eligible layers according to the current allocation gap and page geometry.

\subsection{Batching-Aware Expert Offloading}
\label{sec:design-staging}
Although each token selects only top-$k$ experts, a single batched MoE-layer execution processes the routing decisions of all tokens scheduled in that model step. It may therefore require many more than $k$ distinct experts. Fetching these experts individually after routing would make the MoE kernel wait for the transfers to complete.

For each layer, \sys instead loads the entire contiguous offloaded region into the staging buffer ahead of its MoE execution, regardless of which experts the batch activates.

\section{Implementation}
\label{sec:Imple}

\noindent\textbf{vLLM Integration.}
We implement \sys in the \vllm inference engine with approximately 7,100 lines of code. The implementation integrates value-aware action selection into the \vllm scheduler, implements the dynamic resource orchestrator using CUDA virtual memory management (VMM), and adds batching-aware expert offloading to the MoE execution path. We use PyTorch v2.9.1 and CUDA v12.8.

\noindent\textbf{Cost-Model Parameters.}
The decision policy of \S\ref{sec:design-decision} uses the following configured coefficients: $\alpha{=}0.5$ and $\tau{=}2{,}000$ scheduler steps for the reuse estimate (Eq.~\ref{eq:peff}); and $t_{\text{prefill}}{=}15\,\mu$s and $t_{\text{sched}}{=}500\,\mu$s for the re-prefill and scheduling overheads (Eqs.~\ref{eq:cc},~\ref{eq:cp1}).
The preemption cost initially uses $t_{\text{queue}}{=}1{,}000\,\mu$s and updates this estimate when a preemption-to-rescheduling delay is observed.
The accounting span $w$ in Eq.~\ref{eq:ce} is the minimum remaining decode budget among running requests, clamped to 1--64 decode steps.

We fix the implementation coefficient $c_{\text{reload}}{=}0.63$\,ms in all experiments. With $k{=}10$ and $E{=}512$ rank-local expert shards per layer, this corresponds to $c_{\text{eff}}{=}12.3\,\mu$s in Eq.~\ref{eq:ce}. These values are configured action-comparison weights rather than measured transfer times, and runtime staging transfers the entire offloaded region regardless of the router's realized selections.
We set these coefficients once during development and fixed them across all reported workloads and both TP2 and TP4, without per-workload or per-platform retuning.

\subsection{Implementation of Expert--KV Repartitioning}
\sys remaps physical HBM pages at expert-group granularity. At initialization, the runtime forms each group from the smallest number of consecutive rank-local expert shards whose combined size is aligned to the CUDA VMM allocation granularity, allowing reclaimed pages to be reassigned directly to the KV-cache pool. For Qwen3-Next, each group occupies three 2-MiB pages (6\,MiB): two 3-MiB shards under TP2 and four 1.5-MiB shards under TP4. In all experiments, the controller expands the KV-cache pool in increments of four such groups (\S\ref{sec:design-decision}).

Each TP rank computes how many KV-cache blocks it can add by offloading expert groups. The engine takes the minimum across ranks and instructs every rank to remap that common amount, keeping their KV-cache capacities consistent. The KV allocator inserts the reclaimed pages into its free-block pool, making them immediately available for subsequent allocation. The remapping primitive also supports the reverse operation, but the current controller does not invoke it to restore residency when KV-cache pressure subsides (\S\ref{sec:discussion-limits}).

\subsection{Implementation of Expert Offloading}
\figHidingCopy
\noindent\textbf{Double-Buffered Staging}
 As illustrated in Figure~\ref{fig:hiding_copy}, \sys maintains two staging buffers and alternates them across consecutive layers. While layer $N$ is executing, the runtime asynchronously copies the offloaded weights of layer $N+1$ into the next buffer on a dedicated copy stream.
These copies overlap with the current layer's MoE computation and the dense and router operations preceding the next layer's MoE kernel. The next MoE kernel synchronizes only if its required transfer has not completed. Under TP2, \sys starts the fused gate-and-up GEMM as soon as the corresponding weights have been copied while continuing to transfer the down-projection weights; under TP4, it waits for both weight components before starting the MoE computation.

\noindent\textbf{Boundary-Based Weight Access and Pinned Host Copies}
The evaluated configuration keeps the first expert layer fully resident because the cross-layer prefetch schedule has no preceding expert-layer computation with which to overlap its transfer. For each remaining layer, the residency boundary determines where the MoE kernel reads each expert: it reads lower-index experts from the resident weight tensor and higher-index experts from the staging buffer.

At initialization, \sys keeps the high-index expert weights that may be offloaded from each eligible layer in pinned host memory. Boundary changes determine which of these weights are staged without requiring host-side repacking.

\section{Evaluation}
\label{sec:eval}

\begin{table}[t]
\centering
\caption{Platform configurations.}
\label{tab:platforms}
\small
\setlength{\tabcolsep}{4pt}
\begin{tabular}{lcc}
\toprule
 & \hone (TP=2) & \htwo (TP=4) \\
\midrule
GPUs & 2$\times$H100 PCIe & 4$\times$H200 NVLink \\
HBM per GPU & 94\,GiB & 140\,GiB \\
Expert weights per GPU & 72.0\,GiB & 36.0\,GiB \\
Max.\ batched requests & 128 & 64 \\
KV-cache pool per GPU & 8.02\,GiB & 86.92\,GiB \\
Expert shards per group & 2 & 4 \\
\bottomrule
\end{tabular}
\end{table}
\subsection{Experimental Setup}
\label{sec:eval-setup}

We evaluate how \sys's dynamic HBM repartitioning affects serving performance as KV-cache demand approaches and exceeds the baseline capacity.
Across closed-loop SWE-agent executions, recorded-trace replays, and controlled multi-turn workloads, our experiments collectively examine tail TTFT, TPOT, KV-cache capacity boundaries, and the controller's allocation-time decisions under different expert-offloading ratios.

\paragraph{Model and platforms}
All experiments serve Qwen3-Next-80B-A3B-Instruct, a sparse-MoE model
with 512 routed experts per layer (top-$k{=}10$) plus one shared
expert across 48 layers, 16 query and 2 key/value heads, and a
256K-token native context. We evaluate
on two platforms (Table~\ref{tab:platforms}): a 2$\times$H100 PCIe
testbed, which provides a memory-constrained setting, and a
4$\times$H200 NVLink testbed, which provides larger per-GPU HBM and a
four-GPU tensor-parallel configuration. Both platforms use the same
serving stack, \vllm~0.15.1 with prefix caching and chunked prefill
enabled, to which \sys is applied as a patch set. We refer to the
unmodified configuration as the baseline. Unless otherwise noted,
KV-cache pools are measured at engine start with
\texttt{gpu\_memory\_utilization}${=}0.90$.

We use the two-GPU H100 platform as the primary environment for repeated
experiments under constrained KV-cache capacity. We use the four-GPU
H200 platform for the controlled multi-turn sweep to evaluate whether \sys remains effective with larger HBM, higher tensor parallelism, and a faster interconnect.

\paragraph{Mechanism configuration}
\sys supports both evaluated tensor-parallel configurations (TP=2 and TP=4). We denote by \sys-$r$ a configuration whose expert-offloading ratio is capped at $r\%$ per MoE layer on each GPU. Because
offloading is demand-driven, the realized ratio may remain below this maximum.

\paragraph{Measured KV-cache expansion}
Table~\ref{tab:capacity} reports the HBM that \sys reclaims from the
expert-weight region on \hone: \sys-15 and \sys-20 expand the
8.0\,GiB baseline KV-cache pool to 18.5 and 21.8\,GiB per GPU in our
measurements.

\begin{table}[t]
\centering
\caption{Representative measured KV-cache expansions on \hone{}
(per GPU).
Reclaimed is expert HBM converted to KV-cache backing; ratios are
physical pool sizes, not token capacities.}
\label{tab:capacity}
\small
\begin{tabular}{lccc}
\toprule
Config. & Reclaimed & Total KV pool & KV pool\,/ \\
 & (GiB/GPU) & (GiB/GPU) & baseline \\
\midrule
\vllm   & ---  & 8.0  & 1.00$\times$ \\
\sys-15 & 10.5 & 18.5 & 2.30$\times$ \\
\sys-20 & 13.8 & 21.8 & 2.71$\times$ \\
\bottomrule
\end{tabular}
\end{table}

\paragraph{Workloads}
Table~\ref{tab:workloads} summarizes each workload's role and run
count. The SWE-agent workloads and the LMCache trace run on \hone;
the in-house multi-turn workload runs on both platforms, extending
the sweep to the four-GPU \htwo{} setting. We divide the workloads
into two classes: closed-loop workloads, in which live agents issue
requests based on previous model responses and therefore produce
different request sequences across systems; and replay workloads,
in which every configuration processes the same fixed request
sequence.

\emph{Closed-loop agent execution.}
In this class, live SWE-agent~\cite{yang2024sweagent} executions
attempt a fixed set of 48 SWE-bench Lite tasks~\cite{jimenez2024swebench}
in Docker, selected for their large recorded input-context footprints
(Heavy-48). Because requests depend on prior model responses,
trajectories diverge across systems; we therefore report three runs
each for the baseline and \sys-15 to characterize recurring serving
regimes, not matched-input effects.

\emph{Fixed request sequences.}
The remaining workloads fix their request sequences in advance, so
every configuration processes identical requests under the same
protocol. The main replay re-issues 2{,}103 turns recorded from
SWE-agent executions on the baseline;
replaying this trace under the baseline and \sys-10/15/20 provides
both a system comparison on identical requests and a comparison
across expert-offloading ratios.
The LMCache trace~\cite{lmcache} provides a second recorded workload
for examining the latency--throughput tradeoff across the baseline,
\sys-15, and \sys-20. Each run replays the trace over a fixed
60-minute window, preserving the recorded inter-request think times,
at either 64 or 128 sessions.
The in-house multi-turn workload generates $c$ concurrent synthetic
sessions with deterministic per-turn context growth, allowing us to
vary the reusable prefix-cache footprint and locate the \emph{capacity
boundary}---the load at which this footprint exceeds KV-cache
capacity and TTFT rises sharply (\S\ref{sec:eval-w2}).
\begin{table}[t]
\centering
\caption{Workloads, their evaluation roles, and the number of
independent runs per configuration ($n$).}
\label{tab:workloads}
\footnotesize
\setlength{\tabcolsep}{3.5pt}
\begin{tabular}{@{}p{2.8cm}p{5cm}c@{}}
\toprule
Workload & Role & $n$ \\
\midrule
SWE-agent Heavy-48
& Closed-loop evaluation & 3 \\
SWE-agent replay
& System comparison, identical requests & 5 \\
LMCache trace & Latency--throughput tradeoff & 3 \\
In-house multi-turn
& Capacity-boundary sweep & 5 \\
\bottomrule
\end{tabular}
\end{table}

\paragraph{Metrics and measurement}
We report tail TTFT, TPOT, request throughput for the SWE-agent
replay, and token throughput in tokens per second (TPS). Request throughput is the number
of completed requests within the measurement window divided by its
duration, with each request corresponding to one recorded agent turn,
whereas TPS counts output tokens per second. We also report
prefix-cache hit rate and preemptions as supporting measurements of
cache reuse and memory pressure. Results are computed per run and
reported as mean$\pm$standard deviation (SD) unless
noted otherwise. TTFT uses client first-token timestamps; TPS, cache, and
preemption metrics use server counters; and TPOT is the mean of the
server-side TPOT histogram. For the recorded replay, TPOT is computed
over the full run rather than the measurement interval defined below.
The service-level-objective (SLO) analysis of \S\ref{sec:eval-guidance} instead uses client-measured
TPOT for each request.

\paragraph{Measurement intervals}
For the recorded replay, each run's measurement interval begins after
128 turns have completed and ends when the first session completes
after all 48 sessions have started, before the workload enters its
drain phase; TTFT, request throughput, and preemption rate are computed
over the requests that arrive within this interval. For the in-house
sweep, TTFT percentiles exclude each session's first turn, which is
issued at $t{=}0$ and cannot be served from any prefix cache.

\subsection{Multi-Turn Serving Performance}
\label{sec:eval-stability}
\label{sec:eval-docker}

We combine closed-loop SWE-agent runs with recorded replays: the
former show behavior under agent feedback, while the latter isolate
serving-system differences under fixed request sequences.

\figDockerRegime
\paragraph{Closed-loop execution}
Both systems leave tasks unfinished at the six-hour cap, so completion counts do not provide a controlled comparison. Over the full six hours, the baseline runs accumulate $7{,}227\pm495$ preemptions per run and end with a cumulative prefix-cache hit rate below 0.2\% ($0.105\%\pm0.009\%$).
Over each run's final 30 minutes (5.5--6.0 hours after driver start),
the two systems' observed serving regimes differ sharply. The
baseline averages $0.967\pm0.007$ KV-cache utilization,
$5.05\pm0.52$ waiting requests, $635\pm44$ preemptions, and
$142.5\pm16.5$\,s mean TTFT. \sys-15 averages $0.68\pm0.07$
KV-cache utilization, no waiting requests, no preemptions, and
$2.63\pm0.06$\,s mean TTFT. Because closed-loop trajectories differ
across systems, this comparison characterizes observed serving
regimes rather than a matched-input effect.
Figure~\ref{fig:docker-regime} places three system-state measurements
and mean TTFT on the same six-hour axis ($n{=}3$ per system). In the
baseline, KV-cache utilization pins near capacity, preemptions recur
at roughly 500--800 per 30-minute interval, and waiting requests
build up; over the same period, mean TTFT rises from about 14\,s in
the first bin to more than 150\,s by mid-run. \sys-15 mitigates this
cascade: expert offloading keeps KV-cache utilization below
saturation, preemptions are limited to 1--2 warm-up events per run,
the waiting queue remains near zero, and mean TTFT holds near
2.6\,s with cached prefixes retained. The latency separation in the
bottom panel thus coincides with a persistent separation in all
three system-state panels, rather than appearing only in the final
measurement window.
Figure bins are elapsed-time aligned, whereas the statistics above
use each run's final 30 minutes. Preemption is not itself an
optimization objective: the baseline resorts to it when a KV
allocation cannot be satisfied, and each preempted request re-enters
the queue as a waiting request, independently of the TTFT it later
incurs.

\paragraph{Recorded SWE-agent replay}
Table~\ref{tab:replay-stability} reports the primary system
comparison. Across runs that replay the same 2{,}103 recorded turns
under the same protocol, \sys-15 reduces TTFT
p90 from 26.1\,s to 1.10\,s, increases request throughput by 20.7\%,
and reduces preemptions from 562.7 to 0.18 per 1{,}000 turns. The
cache and queue measurements show two sources of the latency
reduction: the prefix-cache hit rate rises from 0.19\% to
89.8\%, so reused prefixes are not recomputed, and mean queue time
falls from 15.1\,s to 21\,ms, while the preemption rate drops to
nearly zero. The gain holds deeper in the tail, with TTFT p99 falling
from 32.7\,s to 3.48\,s.
Figure~\ref{fig:replay-progress} aligns the systems by recorded turn
number over the whole trace. After the first bin, which includes
turn-0 requests from all 48 sessions, \sys-15 holds TTFT p90 between 0.9 and
2.6\,s in every bin while \vllm{} remains at 21--26\,s, showing that
the aggregate gap does not arise from averaging over different phases
of the trace.
This comes at a decode-time cost: TPOT rises
from 170.9\,ms to 224.0\,ms ($+31.1\%$).
\figReplayProgress
Figure~\ref{fig:turnwise-ttft} reports per-turn median TTFT for
\sys-20 on the same recorded trace (\S\ref{sec:eval-guidance}
compares the ratios directly). At turn~0 the two systems behave
almost identically (medians of 11.6 vs.\ 11.8\,s). Over the
following turns, \sys-20 keeps TTFT near 1\,s, whereas the baseline
climbs to roughly 19\,s and stays high. Beyond roughly turn~25,
baseline TTFT decreases as tasks complete and the number of active
sessions falls (bottom panel). The per-turn view
shows sustained high baseline TTFT while many sessions remain active,
consistent with memory pressure from accumulated reusable prefixes.
We omit turns beyond 61, where fewer than 12 sessions remain.
\figTurnwiseTTFT

\paragraph{LMCache trace}
The LMCache trace shows the same latency--throughput tradeoff on a
second recorded workload ($n{=}3$). At 128 sessions, \sys-15 reduces
TTFT p99 from $29.3\pm0.6$\,s to $4.01\pm0.24$\,s and preemptions from
$1{,}354\pm23$ to zero, while TPS falls from $205.2\pm0.7$ to
$115.2\pm0.2$ and TPOT rises from $118.9\pm0.2$ to
$263.2\pm0.4$\,ms. At 128 sessions, \sys-20 provides TTFT p99
comparable to \sys-15, but has higher TPOT and lower token throughput.
At 64 sessions, it lowers TTFT p99 to
$1.30$\,s, compared with $1.57$\,s for \sys-15. Thus, the ratio that
serves best depends on load.
Within the fixed window, the baseline, \sys-15, and \sys-20 complete
approximately 95\%, 72\%, and 67\% of the recorded turns at 128
sessions. The VAMP configurations complete fewer turns because their
higher TPOT accumulates across successive turns within the fixed
window.
TPS is computed only over each run's active serving period, so a
system that finishes its recorded turns before the window ends is not
penalized for the remaining idle time.

\subsection{KV-Cache Capacity Boundaries}
\label{sec:eval-w2}

\S\ref{sec:eval-stability} evaluates selected workload conditions
but does not locate the \emph{capacity boundary}. We vary session
concurrency in the in-house multi-turn workload to identify this
boundary. At concurrency $c$, the workload runs
$c$ sessions, each following a fixed 12-turn structure with
deterministic per-turn context growth (scaled per platform).
Increasing $c$ therefore increases the reusable prefix-cache
footprint that must remain in HBM.
We evaluate $c\in\{16,20,24,28\}$ on \hone{} and
$c\in\{16,20,24,28,32,36\}$ on \htwo{}, with five independent runs
for each concurrency and expert-offloading ratio. TTFT follows the
steady-state rule of \S\ref{sec:eval-setup}, excluding the
synchronized first turn.
Figure~\ref{fig:w2-sweep} reports steady-state TTFT p99 on both
platforms.

\begin{table}[t]
\centering
\caption{Recorded-trace comparison of \vllm{} and \sys-15 over
2{,}103 turns ($n{=}5$; mean$\pm$SD). TPOT uses the full run; the
other rows use the measurement interval (\S\ref{sec:eval-setup}).}
\label{tab:replay-stability}
\footnotesize
\setlength{\tabcolsep}{3pt}
\begin{tabular}{lccr}
\toprule
Metric & \vllm{} & \sys-15 & Difference \\
\midrule
Request throughput (req/s) & $0.596\pm0.007$ & $0.720\pm0.008$ & $+20.7\%$ \\
TTFT p90 (s) & $26.1\pm0.7$ & $1.10\pm0.17$ & $23.6\times$ lower \\
TTFT p99 (s) & $32.7\pm2.7$ & $3.48\pm0.51$ & $9.4\times$ lower \\
Mean queue time (s) & $15.1\pm0.5$ & $0.021\pm0.012$ & $-99.9\%$ \\
Preemptions / 1k turns & $562.7\pm13.3$ & $0.18\pm0.41$ & $-99.9\%$ \\
Prefix-cache hit rate (\%) & $0.19\pm0.03$ & $89.8\pm3.4$ & $+89.6$\,pp \\
TPOT (ms) & $170.9\pm1.8$ & $224.0\pm2.0$ & $+31.1\%$ \\
\bottomrule
\end{tabular}
\end{table}
\paragraph{\hone: the boundary moves with the expert-offloading ratio}
The baseline is already beyond its KV-cache capacity boundary at
$c{=}16$, with TTFT p99 of $22.1{\pm}0.6$\,s, whereas \sys-25 remains
between 3.0 and 3.9\,s through $c{=}28$. The boundary moves with the
offloading ratio: \sys-15 crosses it at $c{=}24$ ($54.7{\pm}2.1$\,s),
\sys-20 at $c{=}28$ ($63.5{\pm}2.1$\,s), and \sys-25 does not cross it
within the evaluated range. At $c{=}24$, where \sys-15 exceeds its
KV-cache capacity boundary, its TTFT p99 is approximately 26\% higher
than the baseline's. At $c{=}28$, similar aggregate TPS (99--109) masks
different cache regimes: the baseline has a prefix-cache hit rate of
approximately 4\% and 35 preemptions per run, compared with 20--72\%
and approximately one preemption per run across \sys-15/20/25.

\figWTwoSweep
\figRatioTradeoff
\paragraph{\htwo: \sys remains effective at four-GPU scale}
On four H200 GPUs with larger per-GPU HBM and NVLink, all
configurations remain within a 9--11\,s steady-state TTFT p99 band
through $c{=}28$. At $c{=}32$, the baseline crosses its capacity
boundary and its TTFT p99 rises to $92.8{\pm}0.3$\,s, while \sys-20
remains at $9.1{\pm}0.3$\,s, a $10.2\times$ reduction. This TTFT
reduction comes with higher server-mean TPOT: 93.0\,ms for \sys-20
versus 80.1\,ms for the baseline. \sys-15 reaches
$28.3{\pm}0.2$\,s, showing that an undersized ratio only partially avoids
the cliff. \sys-20 and \sys-25 remain below their boundaries at
$c{=}32$ and cross them at $c{=}36$. Below capacity,
overprovisioning still adds transfer cost: at $c{=}28$, \sys-25 has
higher TTFT p99 (11.1 vs.\ 8.9\,s) and TPOT (81.2 vs.\ 66.6\,ms) than
the baseline.

Across both platforms, \sys moves the capacity boundary, the onset
of the prefix-cache cliff, to higher concurrency.

\subsection{Runtime Controller Decisions}
\label{sec:eval-decisions}

We inspect \sys-20 decision logs from the \htwo{} sweep at
$c{=}16$ and $c{=}32$.
At $c{=}16$, sufficient free KV-cache capacity prevents the controller
from being invoked. At $c{=}32$, run~1 records 207 expert-offloading
decisions and 116 cache evictions. Among decisions selecting eviction,
the median eviction score is lower than the corresponding expert-offloading score
($C_c{=}44.9$ vs.\ $C_e{=}52.7$\,ms), consistent with the selected
action; these values are controller scores, not measured latencies.
No preemption is selected, so these traces characterize only the
offloading--eviction choice.

\subsection{Balancing TTFT and TPOT}
\label{sec:eval-guidance}

We compare \vllm{} and \sys-10/15/20 on the same recorded trace
($n{=}5$). Figure~\ref{fig:ratio-tradeoff}(a) shows that \sys-15
achieves the lowest mean TTFT p90 while providing request throughput
comparable to \sys-10. Compared with \sys-15, \sys-20 provides
comparable TTFT p90 but has higher TPOT and lower request throughput. Thus, the expert-offloading ratio should be matched to the workload
rather than maximized.

To evaluate joint latency requirements, we define fixed-window
\emph{SLO-qualified request throughput},
$Q(T,P)=N_{\mathrm{qualified}}(T,P)/W$, where a qualified request
completes within the measurement window $W$ and satisfies
$\mathrm{TTFT}\le T$ and $\mathrm{TPOT}\le P$; failed and
unfinished requests count as misses, and attainment uses
client-measured TPOT for each request (\S\ref{sec:eval-setup}).
Figure~\ref{fig:ratio-tradeoff}(b) evaluates a $5\times5$ grid of
$(T,P)$ requirements. No configuration is best throughout: all provide
little qualified throughput at $P{=}150$\,ms, \vllm{} or \sys-10
leads under the tight $P{=}200$\,ms budget, and \sys-15 leads
throughout $P\ge250$\,ms. Because replay uses fixed concurrency, $Q$
does not estimate a maximum sustainable request rate.

\section{Discussion}
\label{sec:discussion}

\subsection{Interconnect Bandwidth and Copy Hiding}

Expert offload interacts with the interconnect topology, not only with
how much memory is removed. Under TP2 the per-shard MoE kernels are
long, which in principle widens the copy--compute overlap window, but
on our PCIe system the two GPUs communicate over the SYS path across
NUMA domains, so weight transfers contend with tensor-parallel traffic
on the same bus; consistent with this, decode-time cost rises with the
amount of expert weight offloaded
(\S\ref{sec:eval-w2}, \S\ref{sec:eval-guidance}). Under TP4,
GPU-to-GPU tensor-parallel traffic uses NVLink, reducing its
contention with host-to-device weight transfers, although shorter
per-shard kernels provide a smaller overlap window. Because these platforms also differ in GPU model, HBM capacity, and
tensor-parallel degree, the comparison does not isolate
interconnect effects. Exposed expert-offloading overhead depends on
host-to-device transfer time and the amount of computation that
overlaps each transfer.

\subsection{Why We Avoid LRU Expert Cache Execution}

\figLayerDemand

An alternative design maintains a GPU-resident expert cache under an
LRU policy. However, the experts required by each token become known
only after routing. If a required expert is absent from GPU memory, it
must be fetched before the corresponding MoE layer can execute, and
any transfer not hidden by preceding computation directly delays that
layer.

Although each token selects only 10 of 512 routed experts, a batched
MoE kernel requires all distinct experts selected by the tokens in
that model step. Across 1{,}012 profiled model steps, the per-layer
mean number of distinct routed experts ranges from 262 to 473 and
reaches 494 in one step (Figure~\ref{fig:layer-demand}). Thus,
per-token sparsity does not necessarily produce a small expert working
set at layer execution time. Skipping the selected experts that are absent from GPU memory would
change the model output.

Before each layer's MoE kernel, \sys instead stages that layer's
entire offloaded region and overlaps the transfer with preceding
computation. This preserves the routing result while avoiding
per-expert cache-miss handling.

\subsection{Limitations and Future Directions}
\label{sec:discussion-limits}
\sys converts expert HBM into KV-cache capacity, but its coverage ends
when the reusable prefix-cache footprint exceeds the resulting capacity.
Host-side KV-cache offloading is orthogonal and could extend coverage
beyond that point, at a different cost: offloading experts requires no
device-to-host copy at decision time, since the weights already have a
host-resident copy, whereas moving cached blocks out and back adds eviction
decisions and restore latency to the request path. Two prototype limits remain: the maximum offloading ratio trades KV-cache headroom
against staging cost, and the controller does not restore expert
residency when pressure subsides.

\section{Related Work}
\label{sec:related}

\noindent\textbf{MoE inference and expert offloading.}
MoE-Lightning~\cite{cao2025moelightning} reduces MoE inference latency
by streaming cold experts from CPU DRAM and pipelining weight transfers
over PCIe; Fiddler~\cite{kamahori2025fiddler},
KTransformers~\cite{chen2025ktransformers}, and
PowerInfer~\cite{song2023powerinfer} instead execute CPU-resident
model components on the CPU, reducing frequent weight transfers. These systems primarily optimize expert
placement, execution, or transfer, rather than arbitrating expert
offloading against prefix-cache eviction and request preemption at
KV-allocation time.
FluxMoE~\cite{liu2026fluxmoe} similarly uses idle expert HBM for
runtime state through expert paging and budget-aware residency
planning. \sys differs at the decision point: when a KV-cache
allocation cannot be satisfied, it arbitrates expert offloading
against prefix-cache eviction and request preemption rather than
adjusting expert residency alone.
CrossPool~\cite{ye2026crosspool} separates FFN weights and KV caches
into distinct GPU memory pools to colocate multiple cold MoE models; \sys instead
repartitions HBM between the two uses within a single serving model.

\noindent\textbf{Memory management in LLM serving.}
\vllm~\cite{kwon2023vllm} provides paged KV-cache management, and
SGLang~\cite{zheng2024sglang} adds prefix-tree reuse.
vAttention~\cite{prabhu2024vattention} and
GMLake~\cite{guo2024gmlake} use CUDA virtual-memory mapping to reduce
fragmentation, and Oneiros~\cite{li2025oneiros} remaps physical pages
from idle model replicas to the KV-cache pool in multi-tenant serving;
\sys applies the same primitive to a single-tenant MoE model,
reclaiming pages from the serving model's own expert weights.
KunServe~\cite{cheng2026kunserve} frees memory under overload by
dropping parameters replicated across GPUs; \sys instead reclaims
sparsely activated expert groups within a single model, without
requiring replicas.
CachedAttention~\cite{gao2024cachedattention},
LMCache~\cite{lmcache}, MemServe~\cite{hu2024memserve}, and
Mooncake~\cite{qin2024mooncake} persist, reuse, or disaggregate KV
caches across host-memory, storage, and serving instances, while
SnapKV~\cite{snapkv} compresses KV entries. \sys instead expands the
GPU-resident KV-cache pool itself, and its controller compares this
expansion against KV eviction and request preemption.
Prism~\cite{prism} uses GPU-memory ballooning to coordinate memory
across co-served models; \sys instead reallocates HBM within a single
MoE model and arbitrates that expansion against KV eviction and
request preemption.

\noindent\textbf{Serving multi-turn agentic workloads.}
Recent systems schedule around the KV-cache footprint of multi-turn
and agentic workloads: KVFlow~\cite{peng2025kvflow} keeps prefixes of
multi-agent workflows warm, Continuum~\cite{li2025continuum} attaches
a time-to-live to KV caches across turns, and
CONCUR~\cite{chen2026concur} and DualPath~\cite{wu2026dualpath}
manage agentic concurrency and storage bandwidth. These systems take
the HBM KV-cache budget as fixed and decide what to keep within it;
\sys changes the budget by reallocating expert HBM, and
complements such scheduling.

\section{Conclusion}
\label{sec:conclusion}

This paper presents \sys, a serving framework for MoE models in multi-turn agentic environments.
\sys breaks the static boundary between expert-weight memory and the KV-cache pool using CUDA VMM page remapping, guided by a three-way cost model that estimates the future penalties of expert offloading, KV-cache eviction, and request preemption.
Across five runs of the 2{,}103-turn recorded-trace replay, \sys-15 reduces TTFT p90 from 26.1\,s to 1.10\,s (23.6$\times$) and increases request throughput by 20.7\% relative to unmodified \vllm{}, while increasing TPOT by 31.1\%.

\section*{Acknowledgment}
This paper is supported in part by SRC JUMP 2.0 Center for Processing
with Intelligent Storage and Memory (PRISM).

\bibliographystyle{IEEEtranS}
\bibliography{references}

\clearpage
\appendix
\section{Underlying SLO-Qualified-Throughput Slices}
\label{sec:appendix-slices}

Figure~\ref{fig:qsurface-facets} reports the per-budget slices of the
evaluated $5\times5$ $Q(T,P)$ grid summarized by the winner map in
Figure~\ref{fig:ratio-tradeoff}(b). Two structural features stand
out. First,
for every TTFT budget up to 15\,s, the baseline contributes almost no
qualified throughput (at most 0.17\,req/s): its TTFT tail alone
disqualifies most requests, independent of the TPOT budget. Second,
each \sys{} configuration turns on at a characteristic TPOT
threshold that reflects its decode-time cost: \sys-10 rises earliest
but plateaus lowest, \sys-15 rises steeply at $P{=}250$\,ms to
0.6--0.72\,req/s, and \sys-20 contributes only from
$P{=}300$\,ms. The single region where the baseline leads appears
only in the $T{=}30$\,s facet at $P{=}200$\,ms (0.43\,req/s), where
the TTFT budget is loose enough to absorb its tail and the tight TPOT
budget penalizes expert offloading. The winner-map boundaries in
Figure~\ref{fig:ratio-tradeoff}(b) are thus sharp rather than
incidental: they trace which of the two latency constraints binds
first for each configuration.

\figQSurfaceFacets

\section{Closed-Loop Execution: Additional Views}
\label{sec:appendix-docker}

Figure~\ref{fig:docker-three-arm} repeats the panels of
Figure~\ref{fig:docker-regime} with \sys-20 added as the mean of its
two protocol-identical runs; a third \sys-20 run used a different
LM-query timeout and is excluded as a protocol variant. \sys-20
tracks \sys-15's mean KV-cache utilization closely, so the larger
offloading ratio buys little additional average KV headroom on this
workload, while one of its two runs transiently reached saturation
with a burst of 211 preemptions around 1--2\,h (diluted to the small
step in the second panel by the $n{=}2$ mean). With two runs these
are observations, not a reproducibility claim; they are consistent
with selecting \sys-15 as the paper's operating point.

\figDockerThreeArm

Figure~\ref{fig:docker-collapse} shows that the baseline's collapse
is also not specific to the runs of
Figure~\ref{fig:docker-regime}. Across three independent baseline
runs recorded independently on evolving stacks, KV-cache usage
rises to a saturation band above 95\% almost immediately and stays
pinned for each run's full duration (panel~a); cumulative
preemptions grow near-linearly and overlap closely across runs---at
$t{=}7.4$\,h, one run's count of 9{,}183 is within 0.7\% of an
earlier run's final count of 9{,}243, and a 12-hour run continues
along the same trajectory to 13{,}767 (panel~b); and the cumulative
prefix-cache hit rate decays from over 10\% at startup to below
0.2\% in every run (panel~c). The collapse is a property of the
workload--system pair---accumulating reusable context meeting a
fixed KV-cache allocation---rather than an artifact of any
particular run.

\figDockerCollapse

\end{document}